\documentclass{aa}

\usepackage{graphicx}
\usepackage{txfonts}
\usepackage[hidelinks]{hyperref}
\usepackage{booktabs}
\usepackage{makecell}
\usepackage{xcolor}
\begin{document}
\setlength{\columnsep}{28pt}

\title{Phosphorus abundances and multi-element co-enrichment in nearby FGK stars}
\titlerunning{Phosphorus abundances in nearby FGK stars}

\author{Dario Esposito\inst{1}
\and Giovanni Covone\inst{2,3,4}
\and Donato Giovannelli\inst{5,6,7,8,9}
\and Paola Manini\inst{10}
\and Christian Magliano\inst{2}
\and Laura Inno \inst{4,15}
\and Luca Cacciapuoti\inst{11}
\and Víctor M. Rivilla\inst{12}
\and Vito Saggese\inst{2,3}
\and Leonardo Testi\inst{13,14}
\and Emanuele Cristiano\inst{5}
\and Luca Tonietti\inst{15,4}
}

\authorrunning{Esposito et al.}

\institute{Scuola Superiore Meridionale, Largo San Marcellino 10, 80138 Naples, Italy
\and Department of Physics ``Ettore Pancini'', University of Naples Federico II, Naples, Italy
\and INFN section of Naples, Via Cinthia 6, 80126 Napoli, Italy
\and INAF, Osservatorio Astronomico di Capodimonte, Salita Moiariello 16, I-80131 Naples, Italy
\and Department of Biology, University of Naples Federico II, Naples, Italy
\and Institute for Marine Biological Resources and Biotechnologies, National Research Council (CNR-IRBIM), Ancona, Italy
\and Earth-Life Science Institute, Tokyo Institute of Technology, Tokyo, Japan
\and Marine Chemistry \& Geochemistry Department, Woods Hole Oceanographic Institution, USA
\and Department of Marine and Coastal Science, Rutgers University, New Brunswick, NJ, USA
\and Department of Chemical Sciences, University of Naples Federico II, via Cinthia 4, I-80126 Naples, Italy
\and European Southern Observatory, Alonso de C\'ordova 3107, Vitacura, Region Metropolitana de Santiago, Chile
\and Centro de Astrobiolog\'ia (CAB), INTA-CSIC, Carretera de Ajalvir km 4, 28850 Torrej\'on de Ardoz, Madrid, Spain
\and Alma Mater Studiorum -- Universit\`a di Bologna, Dipartimento di Fisica e Astronomia ``Augusto Righi'', Via Gobetti 93/2, I-40129 Bologna, Italy
\and INAF--Osservatorio Astrofisico di Arcetri, Largo E.\ Fermi 5, I-50125 Firenze, Italy
\and Department of Science and Technology, University of Naples Parthenope
Block C4, Centro Direzionale di Napoli (CdN), I-80143, Naples, Italy
}

\date{Received -- / Accepted --}

\abstract
{The distribution of bio-essential elements is key to assessing the chemical `fertility' for life in a galactic context. Among these, phosphorus plays a central role in biomolecules, but it is relatively scarce and difficult to measure in stellar spectra.}
{We aim to characterise phosphorus abundances in nearby FGK stars and quantify how P co-varies with other bio-related elements in different Galactic disk populations.}
{We analysed phosphorus abundances for 233 FGK stars in the solar neighbourhood from the Hypatia Catalog. We studied [P/Fe] as a function of [Fe/H] and other elemental abundances, separating thin- and thick-disk stars. We used non-parametric tests and bootstrap resampling to derive median trends, confidence intervals, correlation coefficients, and residual correlations at fixed [Fe/H]. We also investigated molar ratios involving P relevant to planetary mineralogy.}
{We find a [P/Fe]--[Fe/H] trend broadly similar to that of $\alpha$ elements, supporting a dominant origin in core-collapse supernovae, with some deviations that suggest additional nucleosynthetic channels. Phosphorus shows strong positive correlations with several CHNOPS, $\alpha$, and Fe-peak elements, especially in the thin disk, where stars with higher phosphorus abundances are enriched in multiple bio-essential and rock-forming species. These co-enrichment patterns remain significant in residual analyses at fixed [Fe/H], indicating that they are not a trivial by-product of global metallicity. Thin-disk stars also exhibit higher abundances than thick-disk stars for many elements. The molar P/Mg and P/Si ratios of the Sun are representative of the local stellar population, whereas the relatively elevated solar P/O may have favoured phosphorus bioavailability on Earth.}
{Our study provides a systematic and statistically robust view of how P co-varies with other CHNOPS and $\alpha$ elements in the solar neighbourhood. The results reveal a genuine multi-element co-enrichment pattern beyond metallicity effects and support the use of phosphorus as a tracer of the chemical fertility of galactic environments.}

\keywords{Astrobiology -- Solar neighbourhood -- Stellar abundances}

\maketitle

\section{Introduction} \label{sec:intro}

A subset of elements is essential to terrestrial biochemistry; their nucleosynthetic origin and Galactic distribution constrain the chemical inventory available to planet formation and, ultimately, habitability. Consequently, the search for life elsewhere in the Galaxy is closely related to understanding the diffusion and distribution of these elements.

These elements are C, H, N, O, P, and S (collectively known as CHNOPS), a few alkali and alkaline-earth metals (Na, K, Ca, and Mg), and some transition metals (i.e. V, Mn, Fe, Co, Ni, Cu, and Mo) \citep[e.g.][]{cockell2016habitability, giovannelli2023trace,tonietti2024unveiling, Covone2025}. They make up the fundamental chemical framework of all known life. Moreover, CHNOPS also play a fundamental role in the abiotic dynamics of modern Earth, influencing processes from the top of the atmosphere to the core of our planet \citep{krijt2022chemical}. Among them, phosphorus (hereafter P) is undoubtedly the most enigmatic.

Unlike the other major bio-elements, P is virtually absent from Earth's atmosphere and its main reservoir is rocks \citep{walton2023phosphorus}. Its biogeochemical cycle is dominated by acid-base reactions rather than by redox chemistry, although redox reactions are not entirely absent \citep{pasek2014redox}. Phosphorus may also be strongly partitioned into planetary cores \citep{stewart2007sulfur}, so an underabundance of P relative to the solar value may reduce the amount of surface-accessible phosphorus on exoplanets. Because living cells cannot fix it from the atmosphere, P likely constitutes a limiting nutrient for life on Earth \citep{tyrrell1999relative}. In addition to its geochemical scarcity, P is also cosmically rare, being the least abundant among the CHNOPS elements in both the interstellar medium (ISM) and the Solar System \citep{aschenbrenner2024}.

Complementary constraints on the Galactic phosphorus budget come from the ISM and Solar-System reservoirs. Although P remains observationally elusive in the ISM and is often strongly depleted onto dust, it can become detectable in the gas phase under energetic processing \citep{fontani2024observations}. While ISM phosphorus chemistry is beyond the scope of this work, this broader context helps connect stellar abundances to the forms in which phosphorus becomes available in star- and planet-forming material. Solar-System evidence, including phosphorus detected in comets \citep[e.g.][]{altwegg2016prebiotic, rivilla2020alma} and reduced P phases in meteorites \citep{pasek2008rethinking}, further supports the astrobiological relevance of this connection.

Despite its scarcity, P plays essential roles in cell functioning: it contributes to the build-up of nucleic acids (DNA and RNA) \citep{schlesinger2013chapter}, confers stability to cell membranes, plays a role in cell signalling processes, and is involved in cellular energy storage through adenosine triphosphate (ATP) \citep{butusov2013role}. Another enigmatic aspect of phosphorus is its detection. This element is notably challenging to detect in stellar spectra because its spectral lines are inherently weak and accessible only in the near-infrared or ultraviolet, regions covered by a limited number of high-resolution spectrographs \citep{masseron2020phosphorus}. For this reason, only a few works have successfully measured and analysed phosphorus in stars, as reviewed below. Here we propose an analysis based on the measured abundance of P in FGK stars in our Galactic neighbourhood.

Currently, it is not possible to directly measure the bulk mineralogy composition of
exoplanets' surfaces, as opposed to what can be achieved in the Solar System via remote sensing \citep{he2021optical}. The composition of a star itself can be used as proxy to approximate the elemental composition of a planet. Even if  stars in their interiors have enough heat and pressure to support atomic fusion and complexify their composition, the outermost layer of the star, the photosphere, is largely reflective of the cloud in which it has formed \citep{lodders2019solar}.

Planets, asteroids, moons, and other star-orbiting objects are made up of the same materials present in this cloud. Consequently, because they originate from the same place and the same material available, the abundance of the host star might be used as a proxy for the composition of the planets \citep[see e.g.][]{hinkel2024abundances}. \citet{bond2010compositional} analysed the one-to-one relationship between star's elements abundances and planetary chemistry from a theoretical perspective. Their simulations suggested that the difference in the composition of small planets was likely the result of the differences in abundance of their host stars and not of disk processing during formation.
Of course, the connection between stellar and planetary composition should be interpreted with caution, since the final abundance of several relevant elements can be strongly modified by volatility, planetary differentiation, and the distance from the host star at which the planet forms (e.g.\ inside or beyond the ice line). In particular, N and O are typically treated as highly volatile, whereas P and S are more commonly regarded as moderately volatile \citep{Lodders2003,PalmeJones2003,Suer2023}. Stellar abundance ratios therefore provide a useful starting point for modelling planetary composition and mineralogy \citep{Bond2010,HinkelUnterborn2018}, but cannot be translated directly into final surface abundances or bioavailability.

Despite the observational difficulties associated with phosphorus, several recent studies have investigated its stellar abundance. \citet{maas2019phosphorus} identified a trend in [P/Fe] consistent with $\alpha$-element behaviour and proposed a time-evolution model. Similarly, \citet{hinkel2020influence}, using a sample of 100 stars from the Hypatia Catalog, confirmed a weak negative correlation between [P/Fe] and [Fe/H], and noted that Earth and Mars appear P-rich and N-poor relative to nearby stars.

\citet{masseron2020phosphorus}, analysing data from the APOGEE survey, identified 15 P-rich stars (with [P/Fe] $>$ 1.0 dex) likely belonging to the thick disk or inner halo. These stars also showed enhancements in other $\alpha$ elements, including O, Si, Mg, Al, and Ce.

A larger sample of 163 stars was studied by \citet{maas2022galactic}, who reaffirmed the co-evolution of phosphorus with $\alpha$ elements. \citet{brauner2023unveiling} further investigated P-rich giants and confirmed their association with Si- and Al-rich chemical patterns, suggesting these stars form part of a broader stellar population enriched in multiple elements.

Despite the biological relevance of phosphorus, its stellar abundances remain poorly explored compared to other bio-essential elements. This work aims to help fill this gap by providing a systematic analysis of phosphorus in FGK stars of the solar neighbourhood, making P a useful tracer of chemically rich environments in the solar neighbourhood and providing quantitative constraints on its co-enrichment with other bio-essential elements relevant for planetary composition.

This paper is structured as follows. In Section~\ref{sec:sample}, we describe the sample selection and the construction of the control group. Section~\ref{sec:analysis} presents the elemental abundance analysis and discusses the correlations between phosphorus and other key elements. Finally, in Section~\ref{sec:conclusion}, we summarise our main results and outline their implications.

Overall, our study provides a systematic and statistically robust analysis of phosphorus abundances in FGK stars from the Hypatia Catalog that simultaneously links P to CHNOPS co-enrichment, galactic disk populations, and molar ratios relevant to P-bearing minerals composition. Previous works have mostly focused either on the global [P/Fe]–[Fe/H] trend or on small samples of P measurements, without quantifying in a unified way how P co-varies with other bio-essential and $\alpha$ elements, how these patterns differ between the thin and thick disks, and how they map into molar ratios that control planetary mineralogy and the accessibility of P. Here we fill this gap by combining robust statistical tests with a multi-element abundance analysis, showing that P participates in a genuine CHNOPS and $\alpha$ co-enrichment pattern and establishing P as a tracer of the chemical fertility of galactic environments.

\section{Sample selection} \label{sec:sample}

\subsection{P sample}

We used the Hypatia Catalog Database\footnote{\href{https://hypatiacatalog.com/}{https://hypatiacatalog.com/}} \citep{hinkel2014stellar} because of its comprehensive coverage of stellar elemental abundances in the solar neighbourhood, including rare but biologically relevant elements such as phosphorus. Data are renormalised and standardised to allow for a consistent comparison between different literature sources. In Hypatia, abundances are renormalised to the solar abundance scale of \citet{lodders20094}, which we adopt throughout this work. The catalogue also provides both chemical and kinematic information, making it especially suitable for this study. It is an amalgamated database including the abundances of $\simeq$ 11,000 stars within 500 pc and all exoplanet host stars regardless of distance. A detailed description is given in \cite{hinkel2017hypatia}, and the most recent update in \cite{hinkel2024abundances}.

For our analysis, we selected all stars with phosphorus measurements in Hypatia. By construction, these stars have at least both [P/H] and [Fe/H] measurements \citep{hinkel2014stellar}. The initial sample contains 283 stars. To maximise accuracy, we then required uncertainties in both [P/H] and [Fe/H] below 0.15 dex, leaving 233 stars. We chose this threshold because it corresponds to the upper quartile of the [Fe/H] and [P/H] uncertainty distributions, excluding only the most uncertain measurements while preserving a sufficiently large sample.

This sample reflects both the characteristics of the Hypatia Catalog and our focus on the solar neighbourhood. Because P measurements require high-quality spectroscopy, the sample is naturally biased towards well-studied, nearby, bright, and metal-rich stars. This is not a limitation for the aims of this work, which specifically target reliable phosphorus measurements in the local Galactic environment. The sample is therefore not complete for the whole Milky Way, but remains appropriate for investigating relative abundance trends among well-characterised nearby stars.

The stars span effective temperatures from 4055 to 6870 K and distances from 11 to 463 pc. Spectrally, the sample includes 54 F-type, 143 G-type, and 33 K-type stars, while 3 stars have no spectral classification. The [Fe/H] range is from -1.76 to 0.37 dex, and [P/H] ranges from -1.87 to 0.47 dex. The stars belong to both the thin and thick disk, according to the kinematic classification of \cite{hinkel2014stellar}: 171 are thin-disk stars, 44 thick-disk stars, and 18 remain unclassified. HD~160617 and HD~211998 are classified as thick-disk objects in Hypatia, but are identified as halo stars in the literature. In the main analysis and figures they are retained under the original Hypatia population labels, while the impact of excluding them from the thick-disk fits is discussed separately in Appendix~\ref{app_A}. Distinguishing between thin- and thick-disk stars is important because their different chemical enrichment histories can affect the distribution of bioessential elements and the likelihood of planet formation. Additionally, 14 stars are exoplanet hosts, all in the thin disk; some host multiple planets, bringing the total number of exoplanets in our sample to 20.

\subsection{Control sample}

To assess whether phosphorus-rich stars differ chemically from stars with similar physical properties, we constructed a control sample from the Hypatia Catalog using a k-nearest neighbours (kNN) algorithm \citep{szeker2020weighted}. Our goal was to match each P star with stars that do not have phosphorus measurements but share similar stellar properties.

We selected four parameters: [Fe/H], as a tracer of overall chemical enrichment and to avoid confounding phosphorus trends with general metallicity effects; effective temperature, because of its relation to spectral type and its effect on line strength; surface gravity, because of its influence on atmospheric pressure and line formation; and distance from the Sun, to match the local observational footprint. These features were standardised to avoid scale effects in the distance calculation. For each P star, we then identified the k-nearest neighbours among stars without phosphorus measurements. The optimal value of k was chosen with an elbow-like criterion, selecting the point beyond which larger k did not significantly reduce the mean matching distance. This yielded an optimal choice of k = 20. After removing duplicates, the final control sample contains 2,570 stars.

\begin{figure}
  \centering
  \includegraphics[width=\columnwidth]{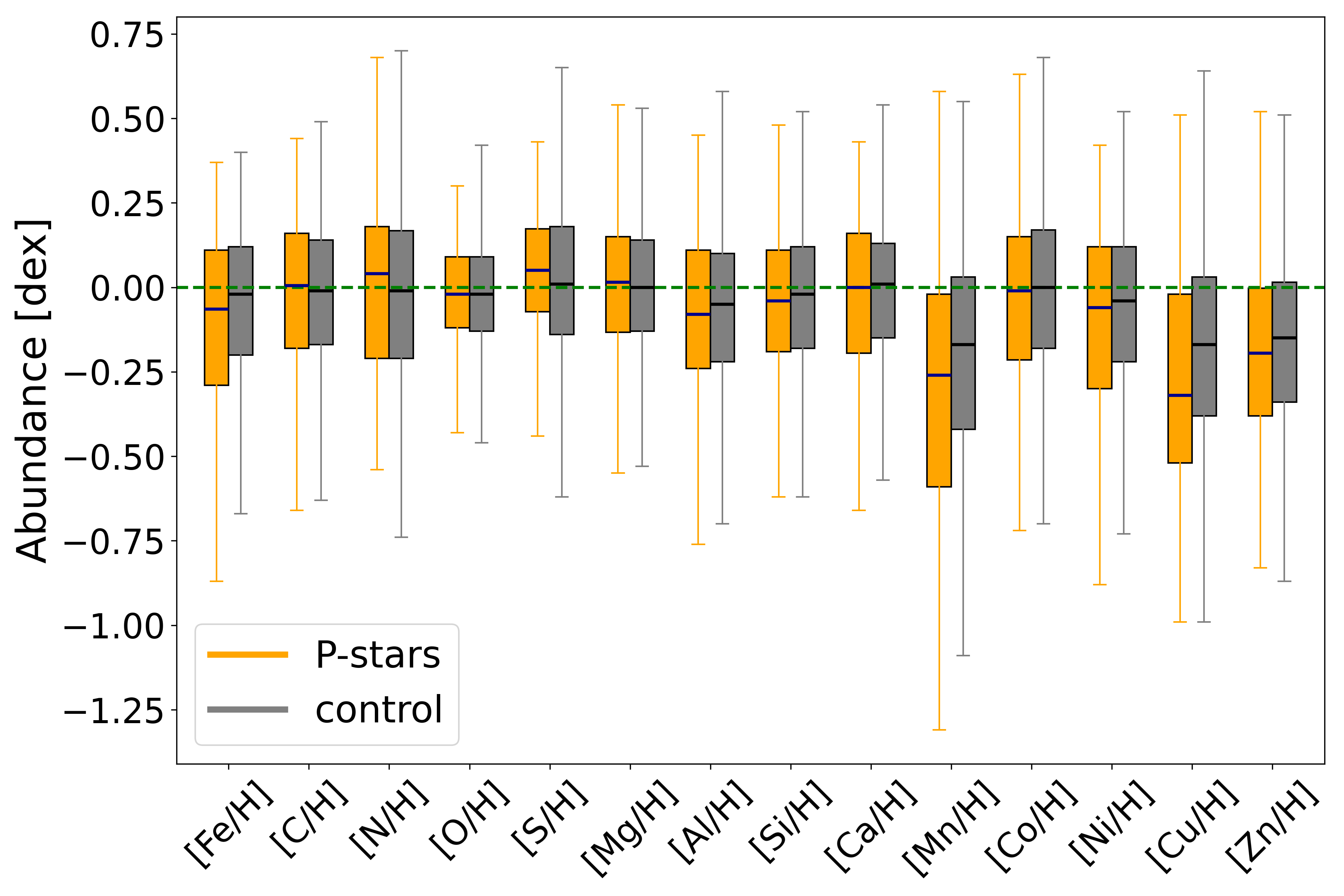}
  \caption{Box plots for the distribution of all life-related elements available for both the sample with phosphorus measurements (indicated as P stars) and the one without them (indicated as control). The abundance distributions are consistent with the absence of strong selection biases between the P star and control samples.}
  \label{fig:box}
\end{figure}

The resulting sample reproduces the distribution of the stellar properties of the P sample and reduces biases related to differences in physical parameters. Finally, we compared the distributions of the selected stellar parameters in the P sample and in the control sample using both the Kolmogorov--Smirnov (KS) and Mann--Whitney U tests. Both confirm that the two samples are statistically consistent in all matched quantities (Table~\ref{table:testcontrol}), supporting the effectiveness of the kNN-based control selection. Figure~\ref{fig:box} shows the corresponding abundance distributions for several life-related elements.

It is important to note that in the Hypatia Catalog the absence of a phosphorus abundance does not distinguish between a non-detection and a measurement not attempted or not reported in the original literature. The control sample should therefore not be interpreted as a population of P-poor or P-free stars, but simply as a parameter-matched reference sample without reported [P/H] measurements, used only to assess selection effects and compare the abundances of other elements.

\begin{table*}
\begin{center}
\caption{Comparison of stellar parameter distributions in the P and control samples.}
\label{table:testcontrol}
\begin{tabular}{l l c c c c c c}
\hline
Parameter  & Test               & Statistic            & p value & median (P stars) & median (control) & $\Delta$ median & $\sigma_{\Delta}$ \\
\hline
Teff       & Mann-Whitney U     & $2.8\times10^{5}$    & 0.21    & 5742             & 5779             &  -37            & 26.95             \\
Teff       & Kolmogorov-Smirnov & 0.067               & 0.28    & 5742             & 5779             &  -37            & 26.95             \\
Distance   & Mann-Whitney U     & $3.1\times10^{5}$    & 0.41    & 59.44            & 56.95            &    2.50         &  3.27             \\
Distance   & Kolmogorov-Smirnov & 0.073               & 0.19    & 59.44            & 56.95            &    2.50         &  3.27             \\
{[}Fe/H{]} & Mann-Whitney U     & $2.8\times10^{5}$    & 0.15    &  -0.07           &  -0.03           &  -0.04          &  0.024            \\
{[}Fe/H{]} & Kolmogorov-Smirnov & 0.076               & 0.16    &  -0.07           &  -0.03           &  -0.04          &  0.024            \\
log g      & Mann-Whitney U     & $2.5\times10^{5}$    & 0.61    &   4.20           &   4.21           &  -0.01          &  0.026            \\
log g      & Kolmogorov-Smirnov & 0.085               & 0.13    &   4.20           &   4.21           &  -0.01          &  0.026            \\
\hline
\end{tabular}
\tablefoot{Kolmogorov-Smirnov and Mann--Whitney U tests compare the distributions of effective temperature, distance, [Fe/H], and surface gravity between the P and control samples. We also report the median values in the two samples and the bootstrap uncertainty on their difference. No statistically significant differences are found in the matched stellar-parameter distributions.}
\end{center}
\end{table*}

\section{Analysis} \label{sec:analysis}
\paragraph{Statistical approach.}
All regression lines, medians, and uncertainty bands reported in this section are supported by non-parametric bootstrap resampling. Unless otherwise stated, we performed $10{,}000$ resamples and report two-sided percentile confidence intervals (95\%, i.e. 2.5--97.5th percentiles). For linear trends, we bootstrapped the slope in each resample; for medians, we bootstrapped the location statistic element by element. This approach avoids distributional assumptions and propagates measurement scatter into the quoted intervals.

\paragraph{Multiple testing.}
Because several families of correlation tests are performed in this work, some nominally significant $p$ values could arise by chance. For the main sets of correlations -- namely [P/H] versus [X/H], [X/Fe] versus [P/Fe], and the residual $\Delta[\mathrm{X/H}]$--$\Delta[\mathrm{P/H}]$ relations -- we therefore applied a Benjamini--Hochberg false discovery rate (FDR) correction to each family of $p$ values and considered as robust those with adjusted $q$ values of $q<0.05$. As a consistency check, the strongest correlations (P with C, O, S, Mg, Si, and the $\alpha$-element proxy) remain significant even when applying a single Benjamini--Hochberg correction to the full set of tests used in this paper.

\subsection{Phosphorus and \texorpdfstring{$\alpha$}{alpha} elements}

\begin{figure*}
  \centering
  \includegraphics[width=0.9\textwidth]{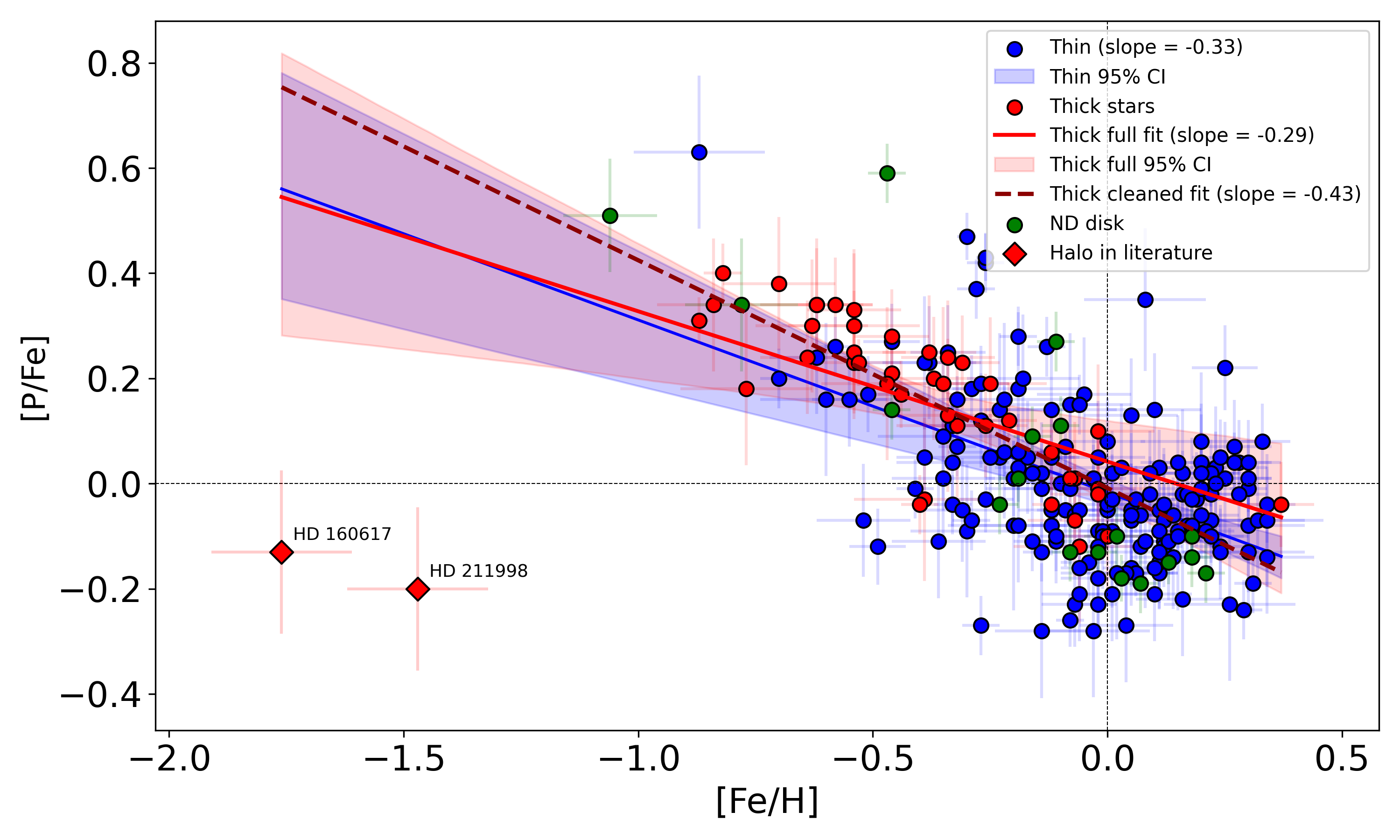}
  \caption{ [P/Fe] as a function of [Fe/H] for the stars in the P sample, separated according to their Galactic component. Blue circles indicate thin-disk stars, red circles thick-disk stars, and green circles stars without a thin- or thick-disk classification in Hypatia. The solid lines show the weighted linear fits for the thin- and thick-disk samples, with the shaded regions indicating the corresponding 95\% bootstrap confidence intervals. HD~160617 and HD~211998 are highlighted with a distinct marker because, although classified as thick-disk objects in Hypatia, they are identified as halo stars in the literature. The dashed red line shows the thick-disk fit obtained after excluding these two stars.}
  \label{fig:P_Fe}
\end{figure*}

As a first step of this analysis, we plot the abundance [P/Fe] of the stars in the P sample as a function of [Fe/H], separating them according to the Galactic component to which they belong; see Figure~\ref{fig:P_Fe}. We find evidence of a weak negative correlation between them for both populations, as found by \cite{hinkel2020influence} and \cite{maas2022galactic}. To quantify the trend, we fitted separate linear relations for thin and thick disks and bootstrapped the slope $10{,}000$ times. The resulting 95\% confidence intervals exclude zero or any positive correlation in both cases, confirming a robust anti-correlation between [P/Fe] and [Fe/H] in each component (see Figure~\ref{fig:P_Fe}). HD~160617 and HD~211998 are highlighted separately in Fig.~\ref{fig:P_Fe}, since they are classified as thick-disk stars in Hypatia but identified as halo stars in the literature; the dashed red line shows the thick-disk fit obtained after excluding these two stars. We further verified that the weighted linear fits provide a reasonable first-order description of the global \([\mathrm{P/Fe}]\)--\([\mathrm{Fe/H}]\) trends, and that the thick-disk relation at low metallicity is sensitive to a small number of objects whose population assignment is uncertain in the literature; the corresponding robustness tests are reported in Appendix \ref{app_A}. Since Hypatia is a heterogeneous compilation of the abundances of the literature, we additionally assessed the robustness of the global \([\mathrm{P/Fe}]\)--\([\mathrm{Fe/H}]\) slope via a bootstrap sensitivity analysis of leave-one-reference-out; the slope remains negative when excluding each individual reference in turn, indicating that our main result is not driven by a single source (see Appendix~\ref{tab:loo_reference_sensitivity}).

\begin{figure*}
  \centering
  \includegraphics[width=0.9\textwidth]{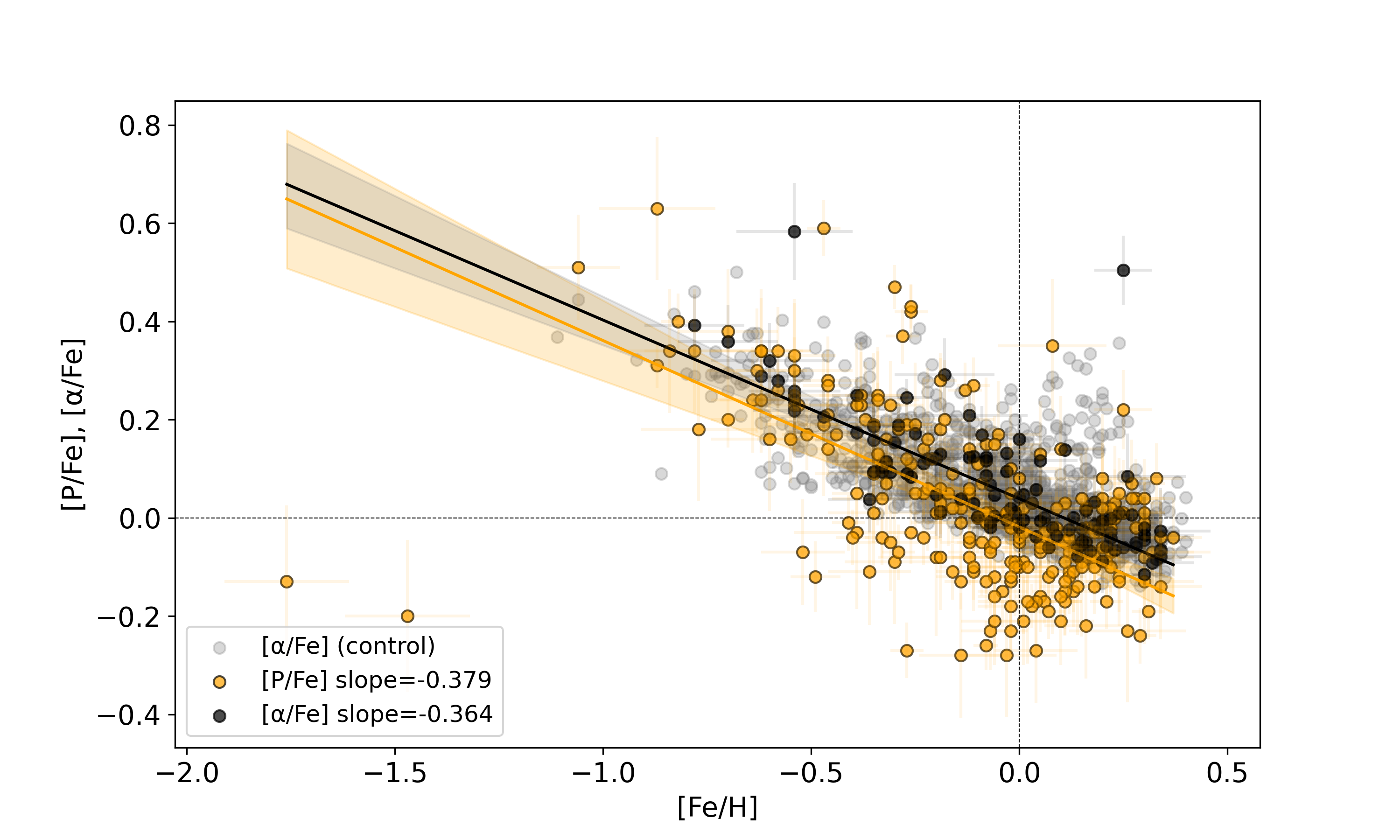}
  \caption{[P/Fe] (orange) and [$\alpha$/Fe] (black) as a function of [Fe/H] for our P sample and the control (in grey, on the background). Solid lines show the best-fit slopes and shaded bands denote 95\% bootstrap confidence intervals on the fit (10,000 resamples). Both ratios anti-correlate with [Fe/H] with comparable slopes ([P/Fe] $\simeq -0.38$, [$\alpha$/Fe] $\simeq -0.36$), but [P/Fe] exhibits a systematically larger scatter and mild offsets relative to [$\alpha$/Fe], indicating that phosphorus does not simply trace classical $\alpha$-element enrichment.}
  \label{fig:alphafe}
\end{figure*}

This behaviour is similar to the [$\alpha$/Fe] trends with respect to [Fe/H], well known in the literature (e.g. \citealt{maas2022galactic}, \citealt{caffau2011galactic}), and shown in Figure~\ref{fig:alphafe} for our sample.
We determined the  [$\alpha$/Fe] abundance ratio for each star with the following method. We selected the available $\alpha$-element measurements for oxygen, magnesium, silicon, sulphur, and calcium. For each element, the abundance relative to iron was calculated. The associated uncertainties in the abundance ratios were propagated using standard error propagation techniques, incorporating the measurement uncertainties of both the element and iron. Then, we calculated the weighted mean of the individual alpha-to-iron abundance ratio to obtain the representative [$\alpha$/Fe] ratio for each star. This approach is similar to the one described in \cite{sanchez2014chemical}, but instead of an arithmetic mean alone we adopted a weighted mean in order to assign greater importance  to elements with smaller uncertainties, ensuring a final value that is more robust and less affected by measurements with higher uncertainty. Additionally, the uncertainty in the weighted mean was derived to reflect the overall reliability of the calculated [$\alpha$/Fe] ratio. For a homogeneous comparison with [P/Fe], the [$\alpha$/Fe] trends versus [Fe/H] were also fitted and evaluated through $10{,}000$ bootstrap resamples, and 95\% confidence bands are shown in Figure \ref{fig:alphafe}.
The same calculation was applied to the control sample (shown in grey in Figure~\ref{fig:alphafe}).
There are 104 stars in the P sample that have measurements for all the $\alpha$ elements.

\begin{figure*}
    \centering
    \includegraphics[width=0.48\textwidth]{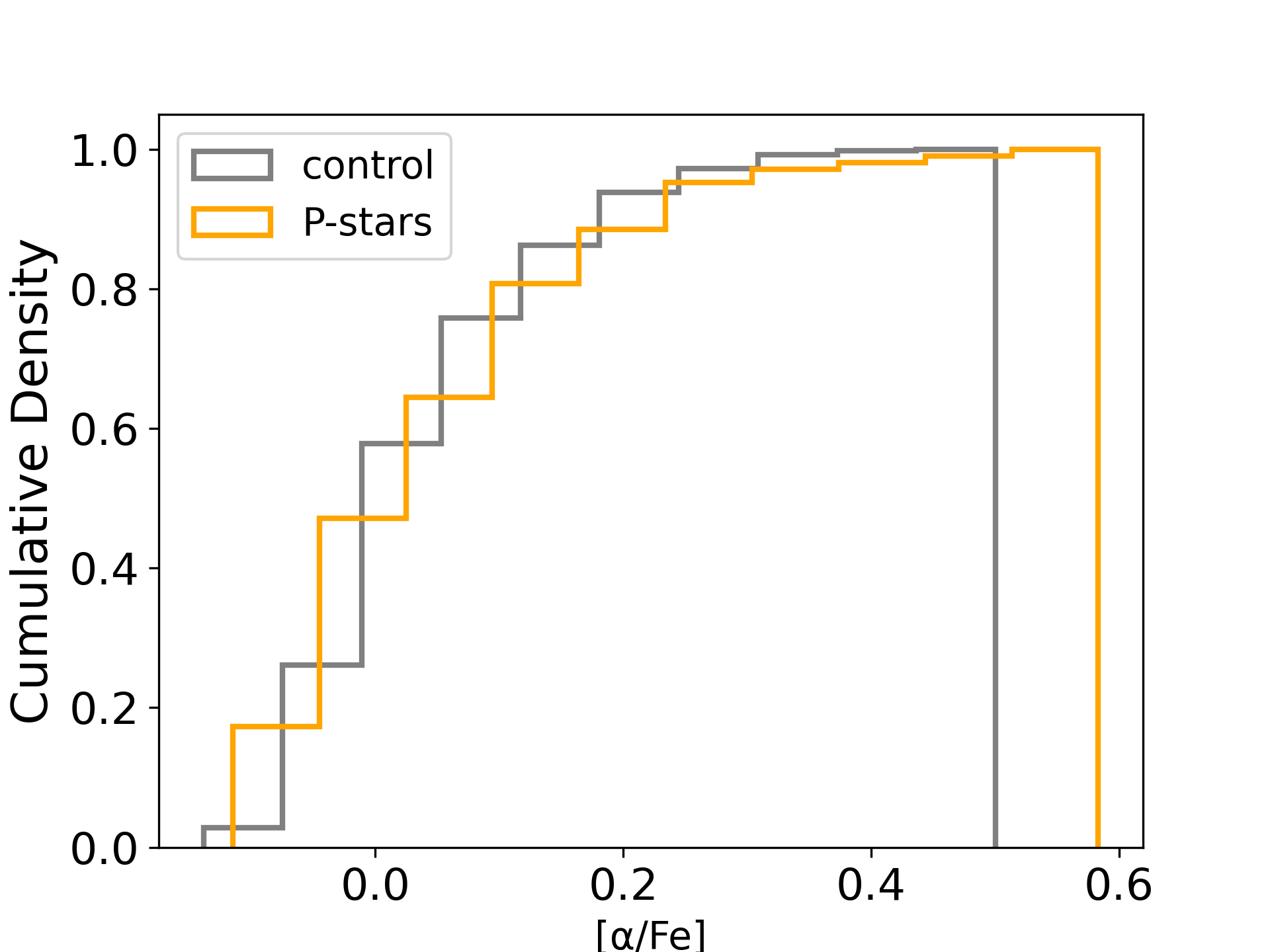}
    \hfill 
    \includegraphics[width=0.48\textwidth]{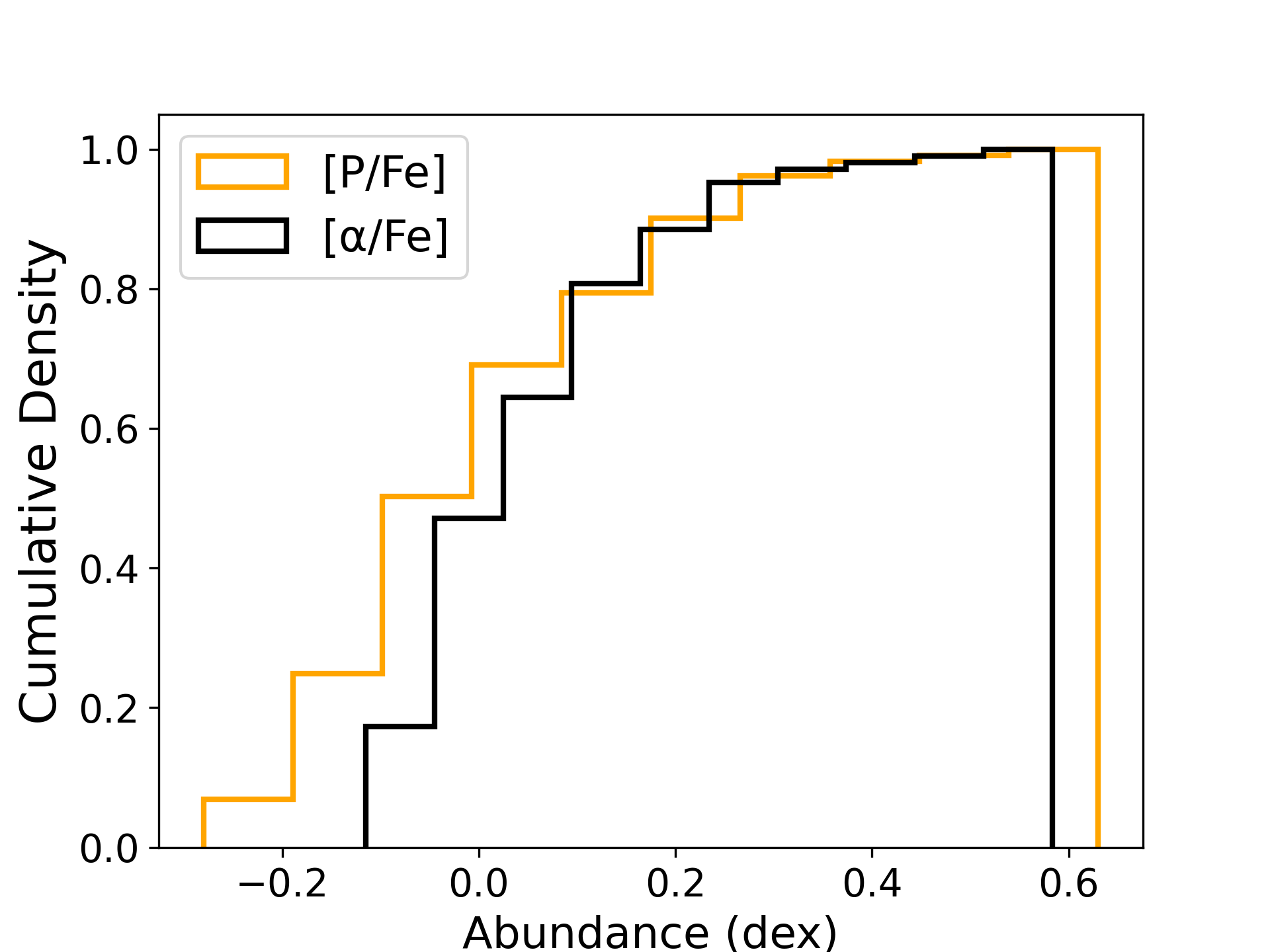}
    \caption{Cumulative distribution of [$\alpha$/Fe] for the control sample and our sample (left panel) and cumulative distribution of [$\alpha$/Fe] and [P/Fe] (right panel). The left panel shows no significant difference between our sample and the control, whereas the right panel reveals that [P/Fe] and [$\alpha$/Fe] follow distinct distributions, indicating that P does not exactly trace classical $\alpha$-element evolution and likely involves additional nucleosynthetic channels.}
    \label{fig:cumuldistr}
\end{figure*}

The cumulative distribution of [$\alpha$/Fe] for our sample and the control is shown in Figure~\ref{fig:cumuldistr} (left panel). We performed a KS test that yielded a KS statistic of 0.1148 and a p value of 0.131. The p value higher than the standard significance threshold (0.05) suggests that there are no significant differences between the two distributions.

Although there is a similar correlation with respect to [Fe/H] for both [P/Fe] and [$\alpha$/Fe], consistent with the co-production of P and $\alpha$ elements (O, Mg, Si, and S), their distributions (see Figure~\ref{fig:cumuldistr}, right panel) exhibit some differences. This is further confirmed by the KS test (KS value = 0.28 and p value = $1.58 \times 10^{-5}$), suggesting that phosphorus does not follow exactly the same chemical evolution as classical $\alpha$ elements. This difference should, however, be interpreted with caution: the broadly similar \([\mathrm{P/Fe}]\) and \([\alpha/\mathrm{Fe}]\) trends with metallicity remain consistent with phosphorus being produced predominantly on short timescales, as expected if core-collapse supernovae provide the dominant contribution. At the same time, the statistically different cumulative distributions indicate that phosphorus does not trace classical $\alpha$-element enrichment in a one-to-one manner, pointing instead to a more complex chemical evolution.

Indeed, $\alpha$ elements are supposed to be principally produced in nucleosynthetic processes of massive stars ($M >$ 8 \(\textup{M}_\odot\)) that culminate their evolution in Type II supernovae. Phosphorus instead, not being a classical $\alpha$ element, forms primarily in Type II supernovae, but also in low-mass (1-3 \(\textup{M}_\odot\)) asymptotic giant branch (AGB) stars by neutron-capture on $^{30}\mathrm{Si}$ \citep{karakas2010updated} and proton-capture on $^{30}\mathrm{Si}$ and $\alpha$ capture on $^{27}\mathrm{Al}$ \citep{caffau2011galactic}. These secondary channels contribute to the Galactic phosphorus budget and may explain the differences observed in its stellar abundance distribution.

\begin{figure*}
  \centering
  \sidecaption
  \includegraphics[width=0.70\textwidth]{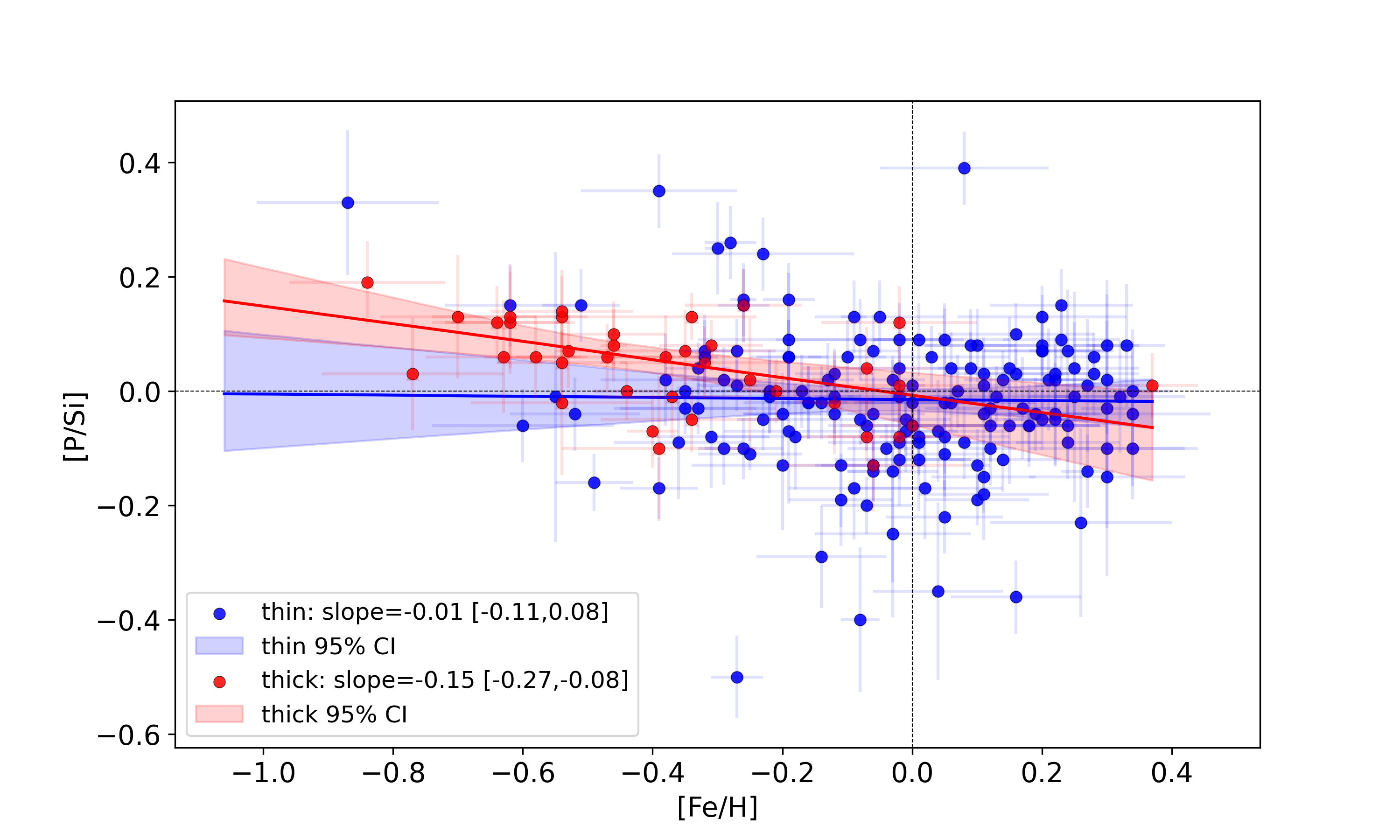}
  \caption{\([\mathrm{P/Si}]\) as a function of metallicity for the thin- and thick-disk subsamples. Thin-disk stars are shown in blue circles and thick-disk stars in red triangles. Solid lines show the weighted bootstrap linear fits (\(10^{4}\) resamples; weights \(w=1/\sigma_{[\mathrm{P/Si}]}\)), with shaded regions indicating the 95\% confidence intervals; slopes are reported in the legend. Two metal-poor outliers with unusually large \([\mathrm{P/Si}]\) uncertainties were excluded from the fit to avoid visual domination, but their removal does not change the inferred slopes within uncertainties. The thin-disk relation is consistent with it being flat, while the thick disk shows a mild negative trend, suggesting that \([\mathrm{P/Si}]\) varies only modestly across the metallicity range probed compared to ratios involving iron.}
  \label{fig:psi_fe}
\end{figure*}

To mitigate the confounding impact of the delayed iron enrichment from Type~Ia supernovae, we also examine the abundance ratio \([\mathrm{P/Si}]\) as a function of metallicity (Fig. \ref{fig:psi_fe}).
Silicon is a canonical \(\alpha\) element, predominantly produced in core-collapse supernovae, while phosphorus is also expected to be dominated by short-timescale enrichment channels, although its yields remain comparatively uncertain. We adopt Si here as a complementary reference element because it is both a major rock-forming species and a convenient Fe-independent comparison for phosphorus. We note, however, that Mg is often regarded as a cleaner tracer of core-collapse enrichment than Si, which can also receive a partial Type~Ia contribution; for this reason, the Si- and Mg-based comparisons should be regarded as complementary rather than interchangeable diagnostics. In the thin disk, the relation is consistent with being flat: the best-fit slope is \(m_{\rm thin}=-0.01\) with a 95\% confidence interval of \([-0.11,\,0.08]\). In the thick disk, instead, we find a mild negative trend, \(m_{\rm thick}=-0.15\) (95\% CI \([-0.27,\,-0.08]\)), corresponding to slightly higher \([\mathrm{P/Si}]\) values at lower \([\mathrm{Fe/H}]\) within this population. A comparison with binned-median trends shows that the weighted linear fits provide an adequate first-order description of the observed behaviour in both disk populations. Overall, across the \(\sim 1.5\)~dex metallicity range probed, \([\mathrm{P/Si}]\) varies only modestly compared to the stronger metallicity dependence typically observed in ratios involving iron (e.g.\ \([\mathrm{P/Fe}]\)), consistent with P broadly tracking a rock-forming \(\alpha\) element over the disk metallicity range sampled. Two metal-poor points with unusually large \([\mathrm{P/Si}]\) uncertainties are shown in the figure but excluded from the fit to avoid visual domination; their removal does not change the inferred slopes within uncertainties.

Interpreting stellar abundances as a proxy for the bulk composition of protoplanetary material, the limited variation in \([\mathrm{P/Si}]\) suggests that the initial phosphorus inventory relative to silicate-forming elements may not differ dramatically across disk metallicities. Under this assumption, rocky planets forming across the sampled metallicity range would start from broadly similar bulk P budgets relative to the silicate matrix, but we acknowledge that subsequent fractionation and differentiation can modify the final mantle and surface reservoirs.

\begin{figure}
  \centering
  \includegraphics[width=\columnwidth]{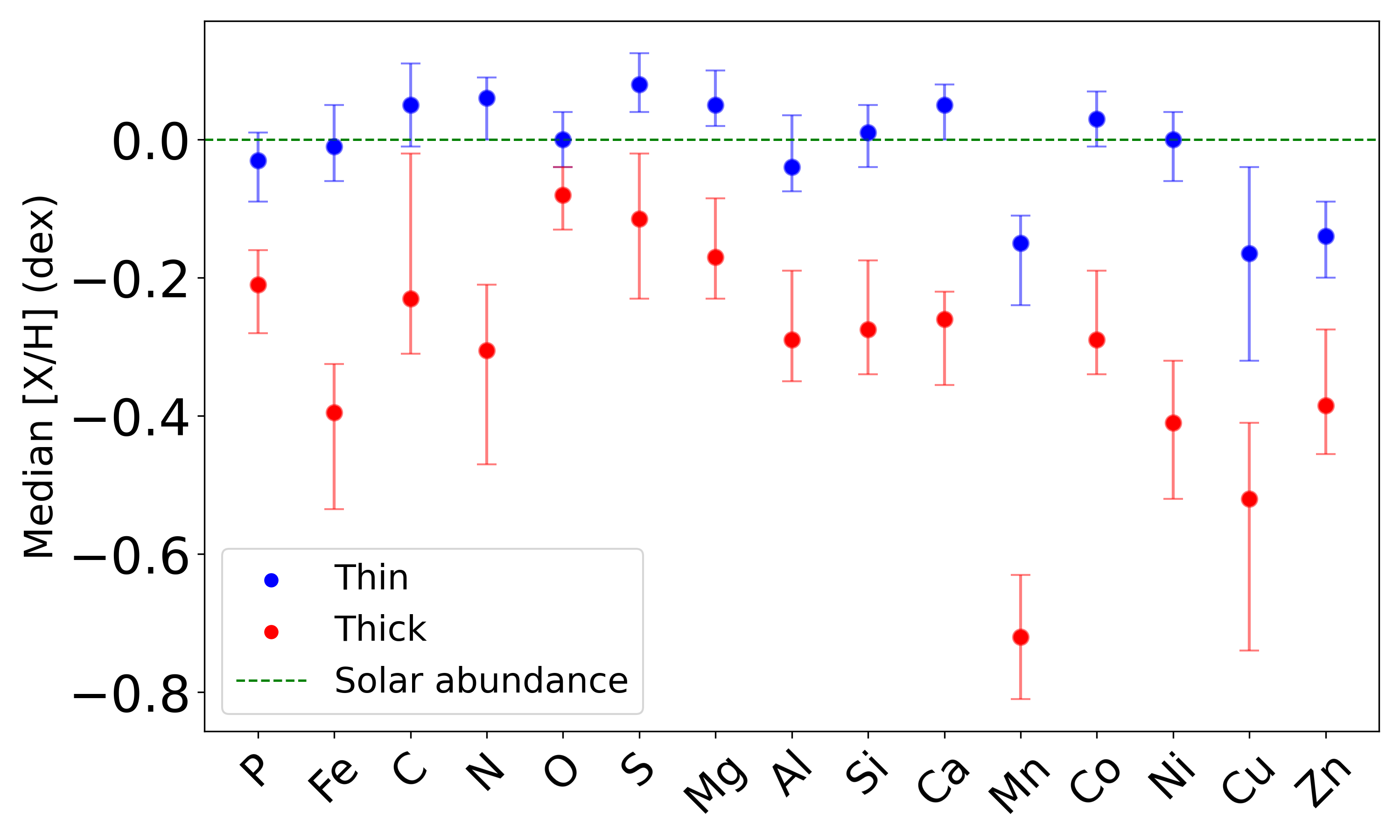}
  \caption{Median elemental abundances \([\mathrm{X}/\mathrm{H}]\) both for the thin (blue) and thick (red) disk samples. Points mark per-element medians (P, Fe, C, N, O, S, Mg, Al, Si, Ca, Mn, Co, Ni, Cu, Zn); error bars are 95\% bootstrap confidence intervals from \(10^{4}\) resamples (percentiles 2.5--97.5). The dashed green line indicates solar \([\mathrm{X}/\mathrm{H}]=0\). Overall, thin-disk medians cluster near solar (several mildly supersolar), whereas thick-disk medians are systematically subsolar by \(\sim 0.2\!-\!0.5\) dex, with the largest offsets for Fe-peak species (e.g. Mn, Ni, Cu) and smaller yet negative shifts for \(\alpha\) and light elements, consistent with an older, more metal-poor thick disk with a distinct enrichment history.}
  \label{fig:med}
\end{figure}

Figure~\ref{fig:med} shows the median abundance of some elements related to life for the thin and the thick disks of our Galaxy. There is a strong difference in the abundance of the majority of the elements taken into account between the thin and thick disk.

To further visualise and quantify these trends, we constructed a set of multi-panel diagrams of [X/Fe]
as a function of [P/Fe] and [Fe/H] for all elements analysed
 (C, N, O, S, Mg, Al, Si, Ca, Mn, Co, Ni,
Cu, and Zn), separately for thin- and thick disk stars
(see Figs.~\ref{fig:xfe_pfe_panels} and
\ref{fig:xfe_fe_panels} in Appendix~\ref{app_B}).
In the thin disk, stars with enhanced [P/Fe] clearly
tend to show enhanced [X/Fe] for most elements,
especially C, O, S, and several $\alpha$ and Fe-peak species, while the thick disk exhibits weaker and more
scattered patterns. The fact that the [X/Fe]--[P/Fe]
relations remain visible even after splitting by disk
population, and are consistent with the residual
correlations at fixed [Fe/H], reinforces the picture of a genuine multi-element co-enrichment with
phosphorus.

We also summarised these trends in correlation tables that report, for each element and disk component, the Spearman coefficients and their bootstrap uncertainties
for [X/Fe] versus [P/Fe] and versus [Fe/H]
(Tables~\ref{tab:xfe_pfe_corr} and \ref{tab:xfe_fe_corr} in Appendix \ref{app_B}).
In the thin disk, the [X/Fe]--[P/Fe] correlations are generally moderate to strong and positive for most elements, with significance levels comparable to those found for the CHNOPS subset, whereas in the thick disk several elements show weaker or poorly
constrained trends.

This dichotomy between the thin and thick disk reflects distinct chemical enrichment histories. The thick disk hosts, on average, older stars that formed earlier, when the ISM had a lower overall metallicity; at fixed [Fe/H], thick-disk stars are typically $\alpha$-enhanced, consistent with shorter enrichment timescales than in the thin disk. Consequently, thick-disk stars generally have lower metallicities than thin-disk stars.

Our results are consistent with previous abundance studies of Galactic disk populations, which have long shown that thick-disk stars are typically more $\alpha$-enhanced and chemically distinct from thin-disk stars at a given metallicity \citep[e.g.][]{adibekyan2012chemical,bensby2014exploring}. In this context, our analysis extends the comparison to phosphorus and to a broader set of bio-essential and rock-forming elements. In particular, we find that the thin-disk population is systematically more enriched in several species, with median offsets of order $\sim 0.2$--$0.5$ dex depending on the element (Fig.~\ref{fig:med}), while the strongest P-linked abundance correlations are observed in the thin disk rather than in the thick disk.

Beyond the global [P/Fe] trend, we also find marked abundance differences between thin- and thick-disk stars. Thin-disk stars are, on average, more enriched than thick-disk stars in several bio-essential (C, N, O, S, P) and rock-forming elements, as well as in many Fe-peak species. Quantitatively, the median abundance offsets between the two populations are typically of the order of \(\sim 0.2\)–0.5 dex, with the largest differences found among Fe-peak species and smaller, though still systematic, deficits for \(\alpha\) and light elements in the thick disk.
This suggests that planets forming around thin-disk stars may inherit a chemically richer inventory of volatile and refractory elements, potentially relevant for planetary mineralogy and for geochemical pathways often discussed in prebiotic scenarios \citep{walton2021phosphorus, li2022minerals, krijt2022chemical}. Conversely, the lower overall enrichment of thick-disk stars points to planetary systems with a more limited chemical budget, even if this does not preclude planet formation itself. Any implications for planetary surface chemistry remain qualitative at this stage and should be tested with dedicated planet-formation and geochemical models. We therefore restrict these conclusions to chemical availability and do not draw quantitative inferences about planetary habitability.

For completeness, we also inspected the small subset of stars in our sample with confirmed exoplanets, comparing their abundances to those of non-host stars. Given the very small number of planet-hosting stars in our sample ($N=14$), and the fact that they are largely dominated by close-in giant planets, we do not attempt any detailed statistical comparison between hosts and non-hosts. This host subsample is strongly affected by well-known detection biases and is not representative of the underlying planet population, especially for potentially habitable, lower-mass planets. We note that planet-hosting stars tend to show somewhat higher abundances for several elements than non-hosts, but given the small and biased nature of this subsample we regard this trend as purely qualitative and do not draw any quantitative conclusions from these data. This cautious interpretation is consistent with previous work by \citet{caffau2011galactic}, who also discussed phosphorus abundances in stars with and without detected planets. In our case, however, the host subsample is substantially smaller and more strongly affected by selection biases, so we limit ourselves to a qualitative comparison and do not attempt a statistically meaningful comparison of host and non-host stars.

\subsection{Phosphorus and CNOS}

\begin{figure*}
    \centering
    \begin{minipage}{0.48\textwidth}
        \centering
        \includegraphics[width=\textwidth]{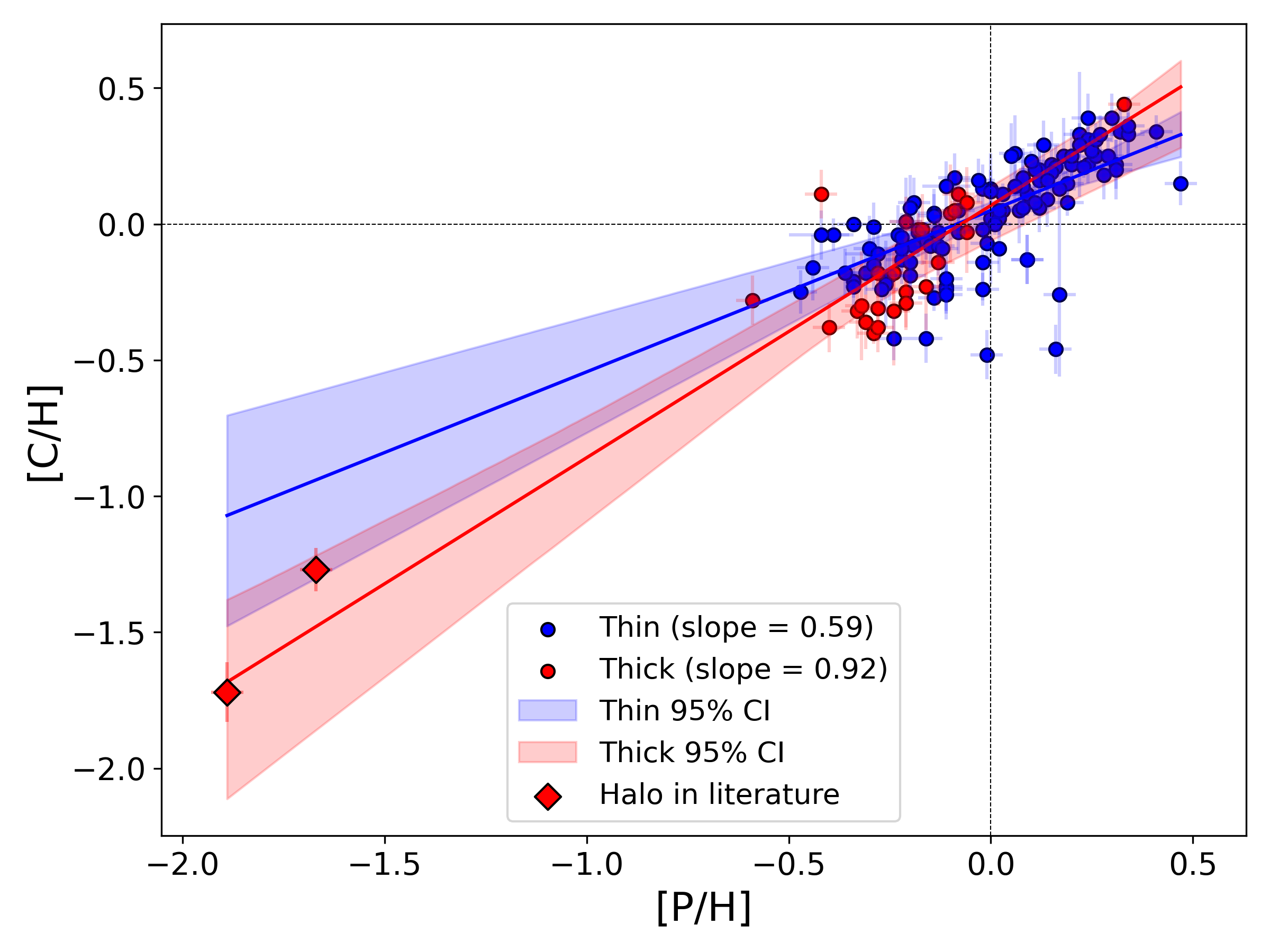}
    \end{minipage} \hfill
    \begin{minipage}{0.48\textwidth}
        \centering
        \includegraphics[width=\textwidth]{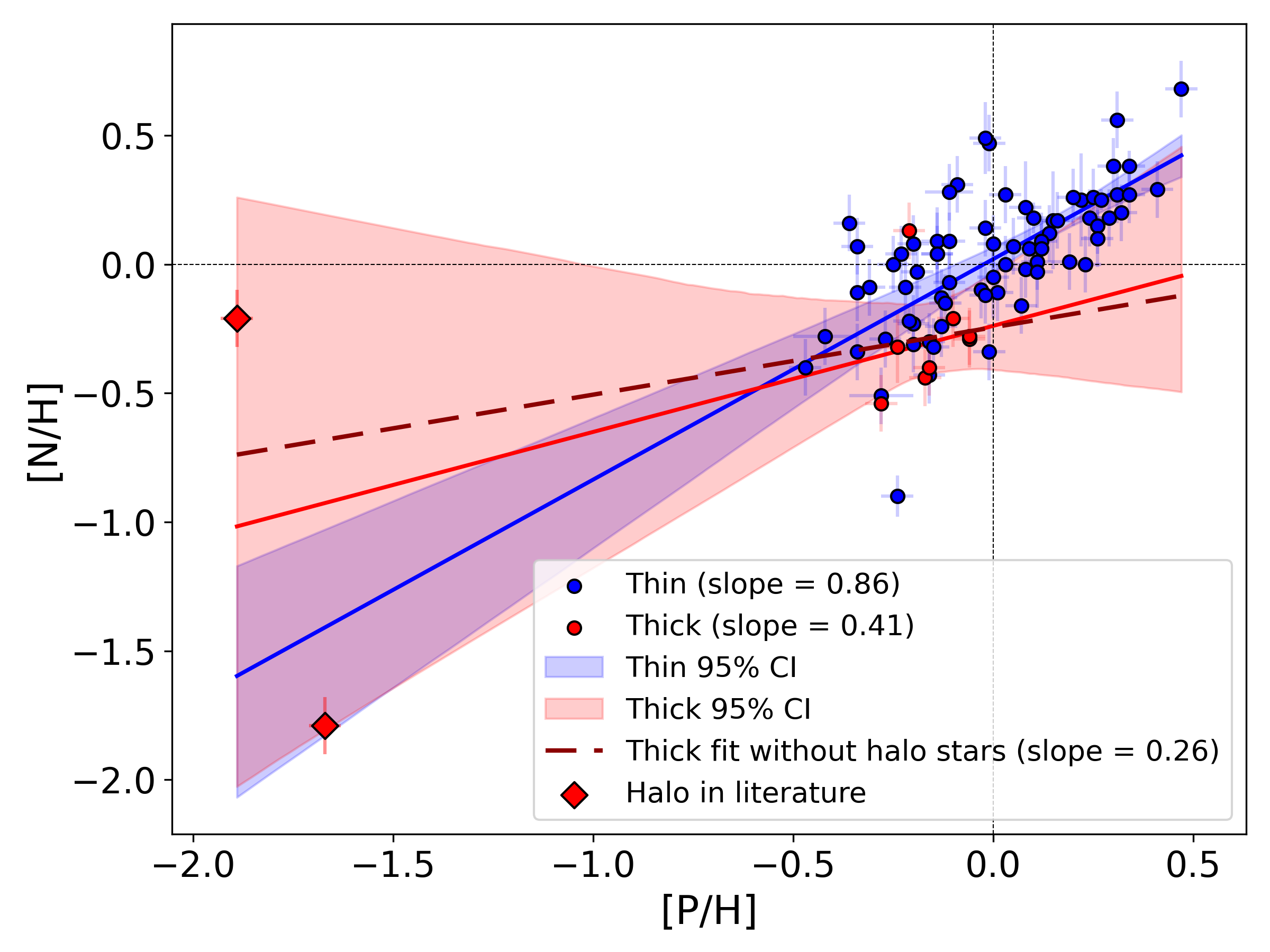}
    \end{minipage}

    \vspace{0.5cm}

    \begin{minipage}{0.48\textwidth}
        \centering
        \includegraphics[width=\textwidth]{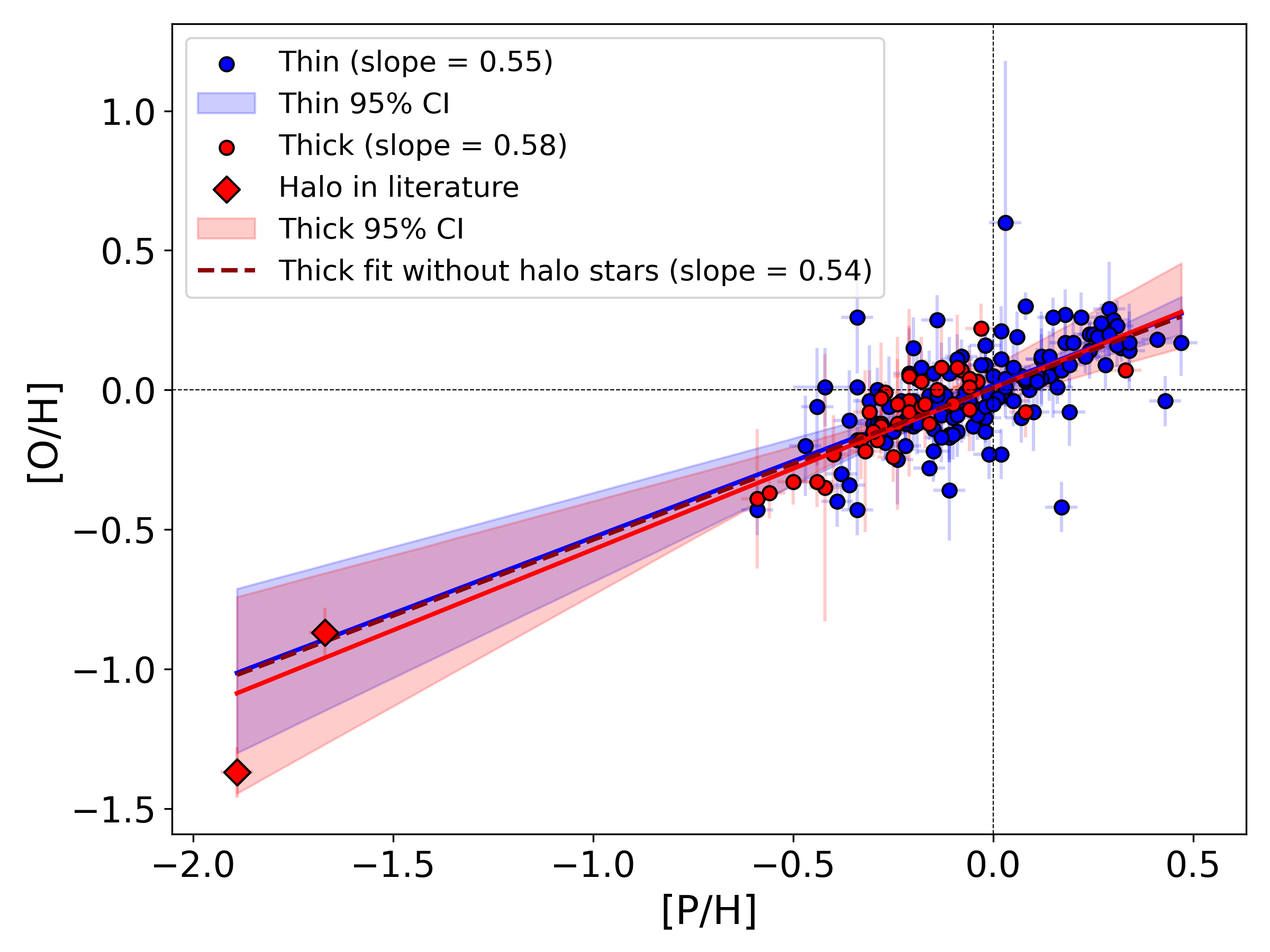}
    \end{minipage} \hfill
    \begin{minipage}{0.48\textwidth}
        \centering
        \includegraphics[width=\textwidth]{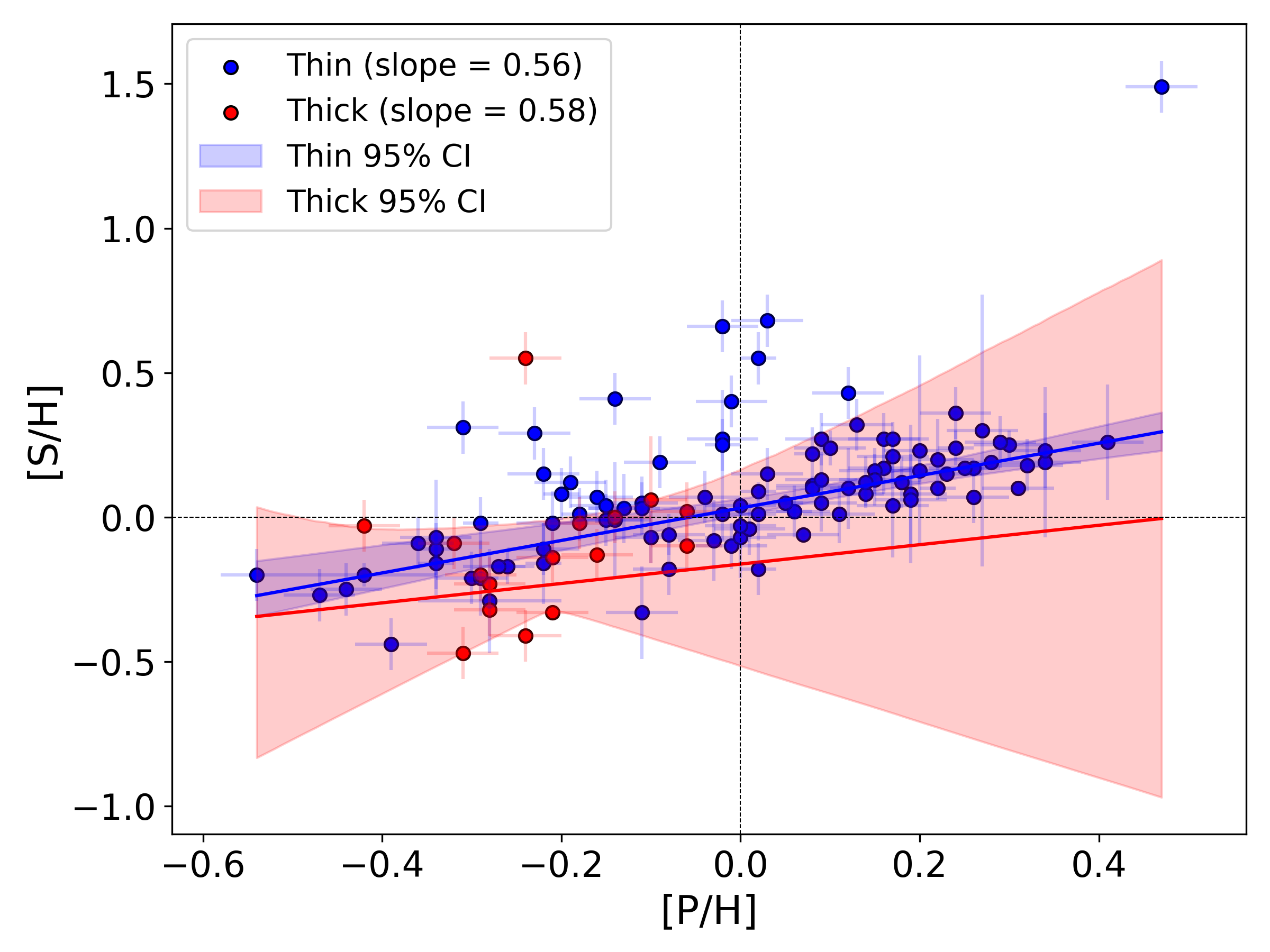}
    \end{minipage}

    \caption{[X/H] versus [P/H] for the stars in the P sample, with X = C, N, O, and S. Blue circles indicate thin-disk stars and red circles thick-disk stars. Solid lines show the weighted linear fits for the thin- and thick-disk samples, with the shaded regions indicating the corresponding 95\% bootstrap confidence intervals. HD~160617 and HD~211998 are highlighted with a distinct marker because, although classified as thick-disk objects in Hypatia, they are identified as halo stars in the literature. Where shown, dashed lines indicate the thick-disk fits obtained after excluding these two stars.}
    \label{fig:cnos}
\end{figure*}

Concerning the most abundant elements in living beings (the CHNOPS) \citep{grigaitis2023makes}, we performed an analysis of them as a function of P. Each analysis includes only stars with both P and the relevant CNOS abundance measured: 144 for C, 83 for N, 176 for O, and 124 for S.
Beyond rank and linear tests, we estimated the slope of [X/H] versus [P/H] via bootstrap to assess uncertainty and sign support. For each element and disk component, $10{,}000$ resamples yield the distribution of slopes; the 95\% intervals and $P(\mathrm{slope}>0)$ are reported in Table \ref{tab:bootstrapcnos} and visualised as shaded bands in Figure \ref{fig:cnos}. HD~160617 and HD~211998 are marked separately in Fig.~\ref{fig:cnos}, since they are classified as thick-disk stars in Hypatia but identified as halo stars in the literature; where shown, dashed lines indicate thick-disk fits obtained after excluding these two stars.
In the thin disk, the bootstrap support for a positive slope is $100\%$ for all CNOS elements, while the thick disk shows broader intervals for N and S.

We further tested whether these correlations are simply driven by the common dependence
of [P/H] and [X/H] on [Fe/H]. For each element X (C, N, O, S, Mg, Si, and the
$\alpha$-element proxy), we fitted [X/H] and [P/H] as functions of [Fe/H] and
computed residuals $\Delta[\mathrm{X/H}]$ and $\Delta[\mathrm{P/H}]$ (see
Table~\ref{tab:residual_corr}). Mg and Si are explicitly included here because they
are key rock-forming elements for terrestrial planets and regulate the formation of the main silicate minerals, including those that incorporate phosphorus; their behaviour is further
discussed in Sect.~3.3. Spearman rank
correlations between these residuals show that P remains moderately correlated with C, O,
Mg, Si, and the $\alpha$-element proxy at fixed [Fe/H] (for C:
$\rho_{\mathrm{S}}\simeq0.42$, $p\simeq2\times10^{-7}$, $N=144$; for O:
$\rho_{\mathrm{S}}\simeq0.44$, $p\simeq9\times10^{-10}$, $N=176$; for Mg:
$\rho_{\mathrm{S}}\simeq0.52$, $p\simeq2.5\times10^{-15}$, $N=203$; for Si:
$\rho_{\mathrm{S}}\simeq0.52$, $p\simeq1.0\times10^{-15}$, $N=203$; for the $\alpha$
proxy: $\rho_{\mathrm{S}}\simeq0.40$, $p\simeq2\times10^{-5}$, $N=104$). Sulphur also
shows a weaker but still significant residual correlation
($\rho_{\mathrm{S}}\simeq0.25$, $p\simeq6\times10^{-3}$, $N=124$), whereas nitrogen does
not show any detectable residual trend ($\rho_{\mathrm{S}}\approx0$, $p\approx0.9$,
$N=83$). After applying an FDR
correction to this set of residual correlations, C, O, S, Mg, Si, and the
$\alpha$-element proxy all remain significant at $q_{\mathrm{FDR}}<0.05$, while
N remains consistent with no correlation. This confirms that, for these
elements, the co-enrichment with P is not merely a reflection of the global
metallicity trend, but traces a genuine pattern of chemical co-variation at
fixed [Fe/H].
Taken together with the trends discussed in Section ~3.1, the residual analysis of
Table~\ref{tab:residual_corr} shows that the P–CHNOPS and P–$\alpha$ correlations are
not merely a by-product of global metallicity, but trace a genuine multi-element co-enrichment pattern that is most clearly seen in the thin disk.

This result is consistent with previous work, but also refines it in several respects. \citet{caffau2011galactic} already showed that phosphorus behaves similarly to the main $\alpha$ elements, while \citet{maas2019phosphorus} and \citet{maas2022galactic} found that the global evolution of P is broadly consistent with that of O, Mg, Si, and S. Likewise, \citet{nandakumar2023m} reported that P follows the main $\alpha$-element patterns in disk stars.
In our sample, the association with P remains significant at fixed metallicity for O, Mg, and Si, with residual Spearman coefficients of $\rho_{\rm S}\simeq 0.44$, $0.52$, and $0.52$, respectively, whereas S shows only a weaker residual trend and N no residual correlation. In addition, the [X/H]--[P/H] bootstrap slopes are uniformly positive and well constrained in the thin disk, while the corresponding thick-disk relations are generally weaker and, in some panels, more sensitive to the inclusion of the two literature-identified halo stars.

The positive correlations observed highlight a fundamental aspect of stellar and planetary chemistry: the environments enriched in P are likely to be a richer inventory of the other elements required for life. This co-enrichment suggests that the abundance of P in a stellar system could serve as an indirect tracer of the overall chemical fertility of its protoplanetary disk. In particular, high abundance of P might indicate places where the building blocks of biomolecules (carbon-based molecules such as lipids, proteins, or nucleic acids) are simultaneously present in sufficient quantity. Our results provide a preliminary indication that planets forming around stars with higher abundances of P may inherit a chemical composition more conducive to prebiotic chemistry. Therefore, identifying regions of the Galaxy where P and other CHNOPS are co-enriched could be a strategic approach to prioritise targets in the search for life elsewhere.

\begin{table}
\centering
\small
\setlength{\tabcolsep}{3pt}
\caption{Bootstrap slopes of [X/H] versus [P/H].}
\label{tab:bootstrapcnos}
\begin{tabular}{llrrrrr}
\hline\hline
Elem. & Sample & Mean & Std & CI low & CI high & $P(>0)$ \\
\hline
C & Thin  & 0.59 & 0.11 & 0.41  & 0.80 & 100\% \\
C & Thick & 0.92 & 0.13 & 0.71  & 1.20 & 100\% \\
N & Thin  & 0.86 & 0.11 & 0.65  & 1.10 & 100\% \\
N & Thick & 0.41 & 0.46 & -0.27 & 1.10 & 74\%  \\
O & Thin  & 0.55 & 0.08 & 0.39  & 0.69 & 100\% \\
O & Thick & 0.58 & 0.11 & 0.38  & 0.81 & 100\% \\
S & Thin  & 0.56 & 0.08 & 0.38  & 0.69 & 100\% \\
S & Thick & 0.34 & 0.67 & -0.96 & 1.70 & 72\%  \\
\hline
\end{tabular}
\tablefoot{For X = C, N, O, and S, we report the mean slope, its standard deviation, the 95\% confidence interval, and the bootstrap probability that the slope is positive, $P(\mathrm{slope}>0)$, separately for the thin and thick disks. Thin-disk stars show uniformly positive, well-constrained slopes, while in the thick disk the trends are generally weaker and less well constrained, especially for N and S. The thick-disk sample follows the original Hypatia classification, and therefore includes HD~160617 and HD~211998, although these stars are identified as halo objects in the literature.}
\end{table}

\begin{table}
\centering
\small
\setlength{\tabcolsep}{6pt}
\caption{Residual correlations at fixed [Fe/H].}
\label{tab:residual_corr}
\begin{tabular}{lcccc}
\hline\hline
Element & $N$ & $\rho_{\mathrm{S}}$ & $p$ value & $q_{\mathrm{FDR}}$ \\
\hline
C          & 144 &  0.42  & $2.2\times10^{-7}$  & $3.8\times10^{-7}$  \\
N          &  83 & -0.01  & $9.0\times10^{-1}$  & $9.0\times10^{-1}$  \\
O          & 176 &  0.44  & $8.9\times10^{-10}$ & $2.1\times10^{-9}$  \\
S          & 124 &  0.25  & $6.0\times10^{-3}$  & $7.0\times10^{-3}$  \\
Mg         & 203 &  0.52  & $2.5\times10^{-15}$ & $8.6\times10^{-15}$ \\
Si         & 203 &  0.52  & $1.0\times10^{-15}$ & $7.1\times10^{-15}$ \\
$\alpha$/Fe & 104 & 0.40  & $2.3\times10^{-5}$  & $3.2\times10^{-5}$  \\
\hline\hline
\end{tabular}
\tablefoot{We report the Spearman rank correlations ($\rho_{\mathrm{S}}$) between residual abundances $\Delta[\mathrm{X/H}]$ and $\Delta[\mathrm{P/H}]$ after subtracting the best-fit [X/H]--[Fe/H] and [P/H]--[Fe/H] trends. The last column gives the Benjamini--Hochberg FDR-adjusted $q_{\mathrm{FDR}}$ values for this family of tests. The residual co-enrichment with P remains moderate and highly significant for C, O, S, Mg, Si, and the $\alpha$-element proxy ($q_{\mathrm{FDR}}<0.05$), while N does not show detectable residual correlation.}
\end{table}

\subsection{Molar ratio} \label{sec:molratio}

In this section, we complement the standard dex notation with molar ratios, since the latter provide a more direct description of stoichiometric and mineralogical relations between elements. This choice follows the approach advocated by \citet{hinkel2020influence} and facilitates the interpretation of stellar abundance patterns in the context of planetary studies.

We calculated the molar ratios using the following equation:

\begin{equation}
\centering
\left( \frac{X}{Y} \right)_\mathrm{mol} =
\frac{10^{[X/H] + \log \epsilon(X)_\odot}}
     {10^{[Y/H] + \log \epsilon(Y)_\odot}}
\label{eq:molar_ratio}
,\end{equation}

where [X/H] and [Y/H] refers to the dex abundance of the elements considered and $\log \epsilon(X)_\odot$ with $\log \epsilon(Y)_\odot$ are the solar reference abundance taken from \citet{lodders2019solar}. This formulation follows Equation (2) from \cite{hinkel2020influence}. This equation converts stellar abundances from  the logarithmic dex scale to molar ratios, suitable for applications in planetary composition and geochemistry. Molar ratios are reported without square brackets, unlike abundance ratios in dex notation, following standard conventions in the literature (\cite{hinkel2020influence}, \citealt{maas2022galactic}).

The classical planet--mineralogy ratios C/O and Mg/Si, which are key diagnostics of the silicate--carbide balance and bulk mantle mineralogy (\citet{larimer1975effect}, \citealt{thiabaud2015elemental}), in our P-star sample are broadly consistent with those of solar neighbourhood FGK stars represented by the control sample, so that our P detections are not restricted to an extreme mineralogical sub-population (see Appendix~\ref{app:CO_MgSi} and Figure~\ref{fig:mgsidistr}).

Phosphorus is relevant to planetary mineralogy because it can occur in both lithophilic and siderophilic phases; in the latter case, it is expected to partition into the iron-rich core during differentiation, reducing the amount of lithophilic P available in the silicate reservoir \citep{todd2022sources}. In this context, we explored the molar ratios P/Si and P/Mg as first-order proxies of the initial bulk chemical inventory available to rocky planets. These two ratios are very tightly correlated in both disk populations, although part of this behaviour is expected algebraically because ratios sharing the same numerator satisfy \(\log(\mathrm{P/Si}) - \log(\mathrm{P/Mg}) = \log(\mathrm{Mg/Si})\). We confirm that the observed scatter is consistent with the Mg/Si distribution (\(\rho_{\rm S} \simeq 1\)), and therefore interpret this diagnostic primarily as a constraint on the initial chemical inventory, rather than on the subsequent differentiation of planetary interiors. A more detailed discussion is given in Appendix~C.2.

\begin{figure*}
\centering
\begin{minipage}[t]{0.62\textwidth}
\centering
\vspace{0pt}
\includegraphics[width=\linewidth]{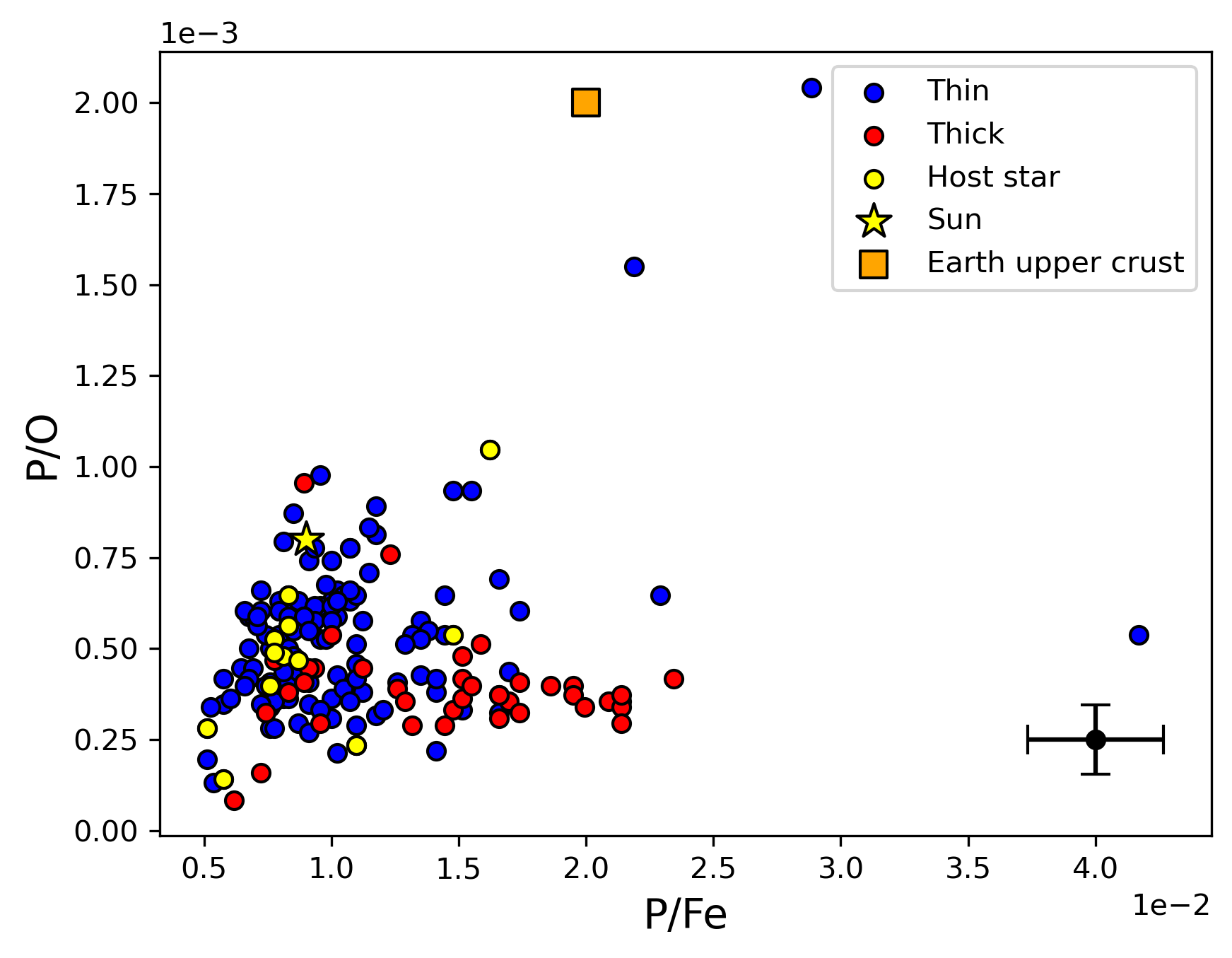}
\end{minipage}\hfill
\begin{minipage}[t]{0.33\textwidth}
\vspace*{0.5cm}
\caption{Scatter plot of P/O ratio in function of the P/Fe. The solar value and Earth crust value are reported with a yellow star and an orange square, respectively. At the bottom right, we show the average uncertainties in P/Fe and P/O. This figure highlights that thin disk stars exhibit a P/O–P/Fe correlation, absent in the thick disk, reflecting different chemical enrichment histories. Moreover, our Sun (indicated with a yellow star) appears lightly enhanced in P/O.}
\label{fig:pfeandpo}
\end{minipage}
\end{figure*}

In an oxidising condition, P tends to form orthophosphate species ($\text{PO}_{4}^{3-}$ ) found in minerals such as apatite, the bioavailable form in Earth crust. Under reducing conditions, instead, phosphorus combines with iron, forming metal-phosphides ($\text{Fe}_3\text{P}$) that segregate into the planetary core during differentiation, reducing the amount of phosphorus accessible to the surface environment and, consequently, to biological processes. To explore these processes from a stellar perspective, we analysed the molar ratios P/Fe and P/O for our sample stars (shown in Figure \ref{fig:pfeandpo}).

This figure presents the distribution of molar P/O as a function of P/Fe for our sample.
Our analysis reveals that thick disk stars exhibit slightly higher P/Fe values compared to thin disk stars, consistent with distinct chemical enrichment for the two populations.

Moreover, the two populations show different correlations between the two ratios: a weak but statistically significant positive
correlation (Spearman rank coefficient $\rho_{\mathrm{S}} \simeq 0.25$,
$p \simeq 3.2 \times 10^{-3}$) is found for the thin disk population,
indicating a statistically significant relationship between the two
variables. On the contrary, no significant correlation is observed
for the thick disk (Spearman $\rho_{\mathrm{S}} \simeq -0.12$,
$p \simeq 4.9 \times 10^{-1}$), suggesting that this trend is not
present in the older stellar population.

The statistically significant correlation in the thin disk, and the absence of correlation in the thick disk, may be interpreted in the context of Galactic chemical evolution. Stars belonging to the thin disk formed in a chemically rich environment in which successive generations of massive stars enriched the ISM with different elements. This progressive enrichment and a more uniform star formation processes may lead to observable correlation between specific elemental abundances. Instead, thick disk stars are formed in an epoch of less efficient chemical mixing and more stochastic enrichment that may obscure any kind of correlation. Thus, this absence of correlation reflects its distinct chemical enrichment history.

The position of the solar P/O ratio, found in the upper range of the densest region of the sample distribution, suggests that the Solar System formed in an environment with a relatively high phosphorus-to-oxygen ratio. This local enrichment could have favoured the formation of abundant phosphate minerals on early Earth, enhancing the bioavailability of P and potentially playing a key role in the emergence of life \citep{hao2025active}. Any such interpretation should, however, be regarded with caution, because the final planetary abundance of several relevant elements can be strongly modified by volatility and differentiation processes, particularly for O and P, which are generally regarded as volatile and moderately volatile, respectively (see Sec. \ref{sec:intro}).

We report the values of P/Fe and P/O for the Earth's crust in the same figure. These ratios are significantly different from stellar values, highlighting the profound impact of planetary differentiation process, such as core formation, mantle-crust segregation and surface oxidation, on the final distribution of P. While stellar abundances set the initial bulk composition of a planet, the accessibility of P is ultimately controlled by planetary differentiation and near-surface processes.

Taken together, these results indicate that P partially co-evolves with $\alpha$ elements, sharing similar nucleosynthetic origins while also being shaped by distinct enrichment processes. Its positive correlations with key biogenic species such as C, O, and S reinforce its role as a tracer of chemically rich environments conducive to biological processes. Overall, our results indicate that the thin and thick disks differ not only in metallicity but also in the relative abundance patterns involving P and other key elements, providing empirical constraints for Galactic chemical evolution and for the initial chemical inventory of planet-forming material.

Finally, the study of molar ratios such as P/Fe and P/O highlights the importance of P in planetary differentiation and surface mineralogy, with potential implications for P bioavailability and the chemical conditions at planetary surfaces. Although stellar abundances provide a valuable starting point for estimating the initial chemical inventory available for biochemical reactions, it is essential to recognise that internal planetary processes, such as core formation and surface oxidation, play a critical role in determining the actual availability of P at the surface. Notably, the Sun exhibits P/Si and P/Mg ratios that are consistent with the bulk of our stellar sample, whereas its P/O ratio appears slightly elevated. This local enrichment in P relative to oxygen could have favoured the formation of phosphate-bearing minerals in the early Solar System, potentially contributing to the favourable chemical conditions that enabled the emergence of life on Earth.

\section{Conclusions} \label{sec:conclusion}

This study explores the Galactic distribution of P and its relationship to other chemical elements relevant to life and planet formation. Using high-quality stellar abundances from the Hypatia Catalog, we analysed a sample of 233 FGK stars with P measurements.

We confirm the similarity between the [P/Fe] and [$\alpha$/Fe] trends with respect to [Fe/H], reinforcing the idea that P is primarily synthesised in massive stars and follows the chemical evolution of $\alpha$ elements, even if the  additional discrepancy in the cumulative distribution between P and $\alpha$ elements suggest that P is produced in Type II supernovae, but also in AGB stars via neutron-capture on $^{30}\mathrm{Si}$ and $\alpha$ capture on $^{27}\mathrm{Al}$.

We further tested a Fe-independent diagnostic by examining \([\mathrm{P/Si}]\) as a function of \([\mathrm{Fe/H}]\). We find that \([\mathrm{P/Si}]\) is consistent with a flat trend in the thin disk, while the thick disk shows only a mild negative slope, implying that \([\mathrm{P/Si}]\) varies modestly across the \(\sim 1.5\)~dex metallicity range probed compared to iron-based ratios.

We find significant differences in the chemical abundances of many elements between the thin and thick disk, not only for P itself, but also for other life-related elements.
Quantitatively, the median abundance differences between thin- and thick-disk stars are typically of the order of $\sim 0.2$--$0.5$ dex, with the largest offsets found among Fe-peak species and smaller, though still systematic, deficits for $\alpha$ and light elements in the thick disk.

This suggests that thin–disk host stars reside in a chemically richer environment, with enhanced availability of several bio-essential elements. In this sense they appear more chemically fertile than thick–disk hosts, but translating these chemical trends into quantitative predictions for the emergence and evolution of life would require additional assumptions and is left to future work.

Among the life-related elements analysed during our study, we paid more attention to CHNOPS, finding a moderate-to-strong positive correlation between the abundance of phosphorus and the abundance of each of the remaining CHNOPS. Quantitatively, the bootstrap support for a positive [X/H]--[P/H] slope in the thin disk is 100\% for all CNOS elements, whereas in the thick disk C and O remain fully supported but N and S are less secure, with positive-slope support decreasing to 74\% and 72\%, respectively. This co-enrichment supports the idea that P-bearing stars may represent favourable sites for the prebiotic reactions and for biochemistry similar to Earth's. These conclusions are robust to sampling noise, as verified by extensive bootstrap resampling of slopes, medians, and correlation coefficients throughout the analysis.

In addition, analysing residual abundances at fixed [Fe/H], we find that P remains significantly correlated with C, O, Mg, Si, and the $\alpha$-element proxy, with residual Spearman coefficients generally $\rho_{\mathrm{S}} \gtrsim 0.4$, while sulphur shows a weaker but still significant residual trend ($\rho_{\mathrm{S}} \sim 0.25$) and nitrogen remains fully consistent with no residual correlation. This confirms that the main P--CHNOPS and P--$\alpha$ correlations we report are not simply a by-product of the underlying [Fe/H] distribution, but carry independent information on the chemical fertility of the star-forming environment.

The analysis of molar ratios shows that the stellar P/Mg and P/Si ratios are very
tightly correlated, consistent with the co-enrichment of P with $\alpha$ elements
such as Mg and Si and with the observed Mg/Si distribution in disk stars. These
trends constrain the initial chemical inventory available to rocky planets, but on
their own do not allow us to infer the details of planet-scale differentiation or
P partitioning.

The correlation for P/Fe versus P/O is low to moderate for the thin disk, while no significant correlation is present for the thick disk. A key point is that the solar P/O value lies in the upper range of the densest region of the distribution. This may indicate a locally enhanced phosphorus abundance compared to average thin-disk stars, which could have influenced the P availability in the early Solar System.

Moreover, the significant difference between the molar ratios observed in the Earth's crust and those derived from stellar abundances highlights how stellar compositions can provide, at best, a modest starting point for estimating the chemical conditions relevant to planetary biogeochemistry. It is therefore essential not to underestimate the critical role of internal planetary processes, such as differentiation and surface geochemistry, in shaping the actual availability of bio-essential elements.

Our work delivers a systematic assessment of phosphorus in FGK stars within Hypatia, linking P to CHNOPS co-enrichment, Galactic disk populations, and molar ratios relevant to planetary mineralogy and phosphorus bioavailability. Altogether, these findings position P as a chemically distinct tracer that partially overlaps with $\alpha$-element evolution, correlates with bio-essential chemistry, and plays a potentially limiting role in planetary habitability. As future stellar surveys (e.g. WEAVE, 4MOST) expand the abundance dataset for under-observed elements such as P, a more comprehensive picture of Galactic habitability will become possible.

\begin{acknowledgements}
DE acknowledges support from the Dottorati PNRR programme, Mission I.4.1 Public Administration, Programme DOT22EBYNH – Molecular Sciences for Earth and Space, CUP F66E23000080006. GC acknowledges support from the University of Napoli Federico II (project:
FRA-CosmoHab, CUP E65F22000050001).
The authors acknowledge the use of the Hypatia Catalog Database, an online compilation of stellar abundance data as described in \citet{hinkel2014stellar}, which was supported by NASA's Nexus for Exoplanet System Science (NExSS) research coordination network and the Vanderbilt Initiative in Data-Intensive Astrophysics (VIDA).
V.M.R. acknowledges support from the grant PID2022-136814NB-I00 by the Spanish Ministry of Science, Innovation and Universities/State Agency of Research MICIU/AEI/10.13039/501100011033 and by ERDF, UE; the grant RYC2020-029387-I funded by MICIU/AEI/10.13039/501100011033 and by "ESF, Investing in your future", and from the Consejo Superior de Investigaciones Cient{\'i}ficas (CSIC) and the Centro de Astrobiolog{\'i}a (CAB) through the project 20225AT015 (Proyectos intramurales especiales del CSIC); and from the grant CNS2023-144464 funded by MICIU/AEI/10.13039/501100011033 and by “European Union NextGenerationEU/PRTR”. DG was supported by funding from the European Research Council (ERC) under the European Union’s Horizon 2020 research and innovation program Grant Agreement No. 948972—COEVOLVE—ERC-2020-STG. This work was partly supported by the Italian Ministero dell'Istruzione, Università e Ricerca through the grant Progetti Premiali 2012 – iALMA (CUP C52I13000140001). This project has received funding from the European Union’s Horizon 2020 research and innovation programme under the European Research Council (ERC) via the ERC Synergy Grant ECOGAL
(grant 855130).
\end{acknowledgements}

\bibliographystyle{aa}
\bibliography{bibliography}

\begin{appendix}
\nolinenumbers
\section{Sensitivity to abundance sources}\label{app_A}

For transparency, we first list in Table~\ref{tab:p_sources} the original literature sources of the phosphorus abundances compiled in the Hypatia Catalog and used in this work, together with the number of stars contributed by each study. This table is intended to document the bibliographic origin of the \([\mathrm{P/H}]\) measurements adopted here, including sources contributing only a small number of stars. We then used the subset of references with sufficient sample size to perform the leave-one-reference-out sensitivity analysis described below.

To assess whether the global \([\mathrm{P/Fe}]\)--\([\mathrm{Fe/H}]\) trend could be driven by a single literature source or by reference-dependent systematics in the Hypatia compilation, we performed a leave-one-reference-out (LOO) sensitivity analysis.
For each reference contributing \([\mathrm{P/Fe}]\) measurements to our sample, we removed all stars associated with that reference and re-estimated the global linear slope using the remaining stars.
For each LOO realisation, we derived the slope distribution using non-parametric bootstrap resampling of the \(([\mathrm{Fe/H}], [\mathrm{P/Fe}])\) pairs and reported the median slope and the 95\% bootstrap confidence interval. The results (Table~\ref{tab:loo_reference_sensitivity}) show that the slope remains negative for all LOO realisations, indicating that the anti-correlation between \([\mathrm{P/Fe}]\) and \([\mathrm{Fe/H}]\) does not depend on any single reference.

As an additional robustness check, we verified that the global \([\mathrm{P/Fe}]\)--\([\mathrm{Fe/H}]\) anti-correlation is not driven by the low-metallicity tail of the sample and that a single weighted linear fit provides a reasonable first-order description of the trend. We therefore compared three thick-disk samples: the full sample, the conservative sample excluding stars with \([\mathrm{Fe/H}] < -0.8\), and a halo-cleaned sample in which only the stars classified as halo in the literature were excluded. Table~\ref{tab:pfe_thick_cleaned} summarises the corresponding slopes, 95\% bootstrap confidence intervals, and the agreement between the weighted linear fits and the corresponding binned median trends. The thick-disk slope changes from \(-0.285\) in the full sample to \(-0.453\) in the conservative sample, and to \(-0.447\) in the halo-cleaned sample. We also cross-checked the most metal-poor stars classified as thick disk in Hypatia against the literature. Among the five objects with \([\mathrm{Fe/H}] < -0.8\), HD~211998 and HD~160617 are explicitly classified as halo stars in the literature \citep{molaro1997new,peterson2020trans}, whereas HD~59374 is more likely associated with the thick disk than with the halo \citep{smiljanic2009beryllium}. No equally clear halo classification was found for HD~113092 or HD~81762. The close agreement between the conservative and halo-cleaned slopes indicates that the difference between the full and conservative thick-disk fits is largely driven by a small number of metal-poor halo contaminants.

\begin{table}[!b]
\caption{Literature sources of the phosphorus abundances used in this work.}
\label{tab:p_sources}
\centering
\small
\begin{tabular}{lcc}
\hline\hline
Reference & Year & $N_{\mathrm{stars}}$ \\
\hline
Maas et al.      & 2022 & 142 \\
Maas et al.      & 2019 & 30  \\
Caffau et al.    & 2019 & 20  \\
Nandakumar et al.& 2022 & 15  \\
Sneden et al.    & 2021 & 13  \\
Jacobson et al.  & 2014 & 5   \\
Maas et al.      & 2017 & 3   \\
Roederer et al.  & 2014b & 2  \\
Caffau et al.    & 2015b & 1  \\
Caffau et al.    & 2011 & 1   \\
\hline
\end{tabular}
\tablefoot{For each study we report the number of stars contributing phosphorus measurements. The corresponding observational setups are: HPF at HET for Maas et al.~(2022) and Sneden et al.~(2021); Phoenix at Gemini South for Maas et al.~(2019); GIANO at TNG for Caffau et al.~(2019); IGRINS for Nandakumar et al.~(2022); HST/STIS near-UV archival data for Jacobson et al.~(2014) and Roederer et al.~(2014b); Phoenix at KPNO Mayall 4 m telescope and Gemini South for Maas et al.~(2017); GIANO at TNG, with CRIRES at VLT for some stars, for Caffau et al.~(2015b); and CRIRES at VLT for Caffau et al.~(2011). Sources contributing only a very small number of stars are listed here for completeness, while the subset included in the leave-one-reference-out sensitivity analysis is reported separately in Table~\ref{tab:loo_reference_sensitivity}.}
\end{table}

\begin{table*}
\centering
\renewcommand{\arraystretch}{1.2}
\caption{Leave-one-reference-out sensitivity of the global \([\mathrm{P/Fe}]\)--\([\mathrm{Fe/H}]\) slope.}
\label{tab:loo_reference_sensitivity}
\begin{tabular}{lccccc}
\hline
\makecell[l]{Excluded\\reference} &
\makecell[c]{$N_{\mathrm{used}}$} &
\makecell[c]{Slope\\median} &
\makecell[c]{Slope\\95\% CI} &
\makecell[c]{$\Delta$Slope} &
\makecell[c]{$\Delta$Slope\\95\% CI} \\
\hline
None (full sample)        & 233 & $-0.29$ & [$-0.42$, $-0.18$] & $+0.00$ & [$+0.00$, $+0.00$] \\
Maas et al.\ (2022)       &  91 & $-0.15$ & [$-0.39$, $-0.02$] & $+0.14$ & [$-0.13$, $+0.33$] \\
Caffau et al.\ (2019)     & 213 & $-0.31$ & [$-0.45$, $-0.18$] & $-0.02$ & [$-0.20$, $+0.16$] \\
Sneden et al.\ (2021)     & 220 & $-0.28$ & [$-0.39$, $-0.16$] & $+0.02$ & [$-0.15$, $+0.18$] \\
Nandakumar et al.\ (2022) & 218 & $-0.28$ & [$-0.41$, $-0.17$] & $+0.01$ & [$-0.16$, $+0.18$] \\
Maas et al.\ (2019)       & 203 & $-0.30$ & [$-0.43$, $-0.17$] & $-0.00$ & [$-0.19$, $+0.17$] \\
\hline
\end{tabular}
\tablefoot{For each iteration we remove one literature reference (as listed in Hypatia for the P measurements), refit a linear trend to the remaining stars, and estimate the slope distribution via non-parametric bootstrap resampling. The table reports the median slope and its 95\% bootstrap confidence interval, together with the change in slope relative to the fit obtained using the full sample. This quantifies the robustness of the inferred trend against potential reference-dependent systematics and sample dominance.}
\end{table*}

\begin{table}
\caption{Thick-disk \([\mathrm{P/Fe}]\)--\([\mathrm{Fe/H}]\) slopes for the full, conservative, and halo-cleaned samples.}
\label{tab:pfe_thick_cleaned}
\centering
\footnotesize
\setlength{\tabcolsep}{3pt}
\begin{tabular}{l c c c c c}
\hline\hline
Sample & $N$ & Slope & 95\% CI & $\langle |\Delta| \rangle$ & $\langle |z| \rangle$ \\
\hline
Full thick    & 44 & $-0.285$ & [$-0.481$,$-0.101$] & 0.043 & 0.443 \\
Cons.\ thick  & 39 & $-0.453$ & [$-0.597$,$-0.339$] & 0.046 & 0.664 \\
Cleaned thick & 42 & $-0.447$ & [$-0.563$,$-0.359$] & 0.018 & 0.936 \\
\hline
\end{tabular}
\tablefoot{The conservative sample excludes stars with \([\mathrm{Fe/H}] < -0.8\), while the halo-cleaned sample excludes HD~211998 and HD~160617, which are classified as halo stars in the literature. $\langle |\Delta| \rangle$ is the mean absolute difference, in dex, between the weighted linear fit and the corresponding binned median trend, and $\langle |z| \rangle$ is the mean absolute normalised residual.}
\end{table}

\section{Multi-element abundance panels and correlation tables}\label{app_B}

\begin{figure*}[p]
    \centering
    \includegraphics[height=0.9\textheight]{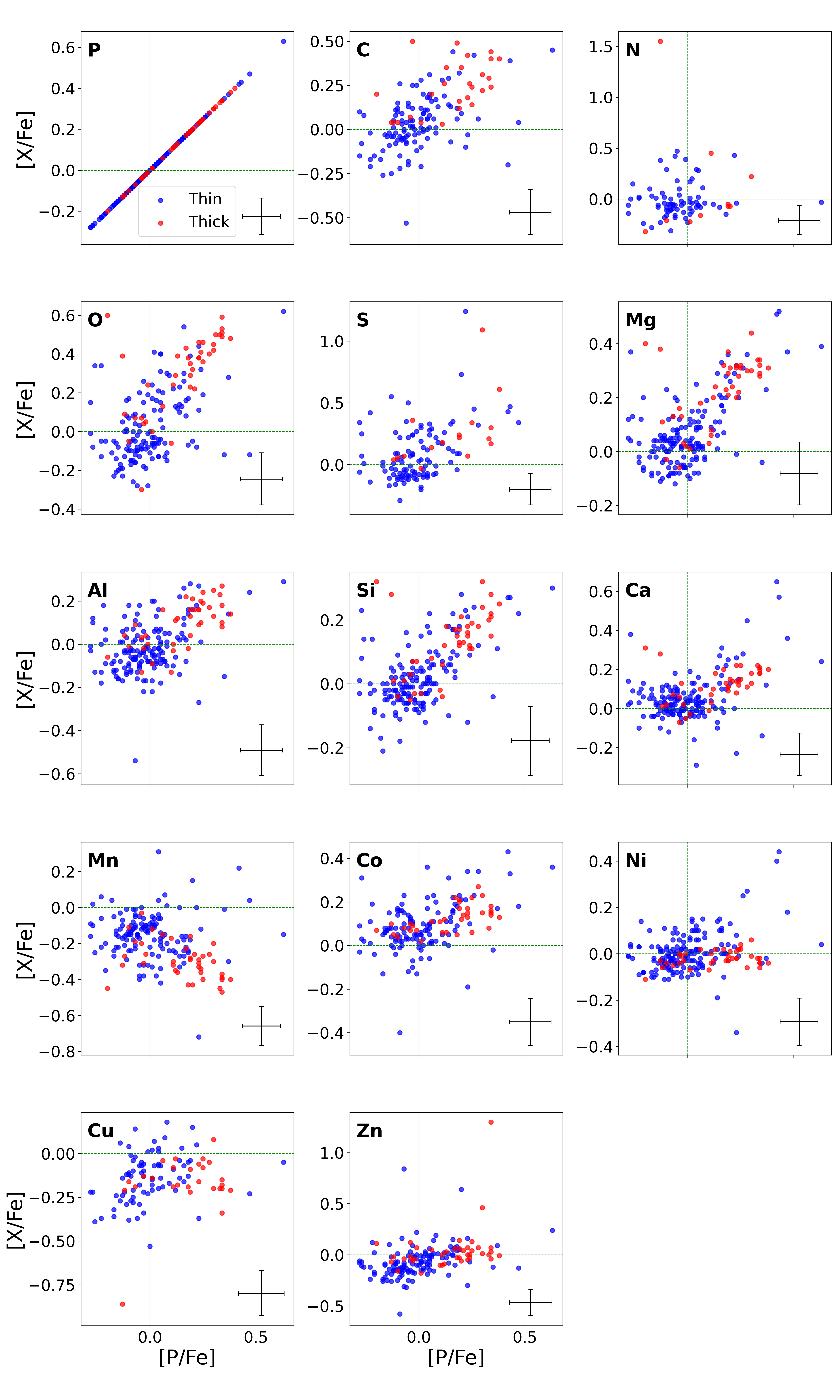}
    \caption{
    Multi-panel diagram of [X/Fe] versus [P/Fe] for all
    elements analysed (C, N, O, S, Mg, Al, Si, Ca, Mn,
    Co, Ni, Cu, Zn). Blue and red points correspond to
    thin- and thick-disk stars, respectively; dashed lines
    mark the solar ratios ([P/Fe] = 0, [X/Fe] = 0). A typical
    1$\sigma$ error bar in [P/Fe] and [X/Fe] is shown in the
    lower-right corner of each panel.
    }
    \label{fig:xfe_pfe_panels}
\end{figure*}

\begin{figure*}[p]
    \centering
    \includegraphics[height=0.9\textheight]{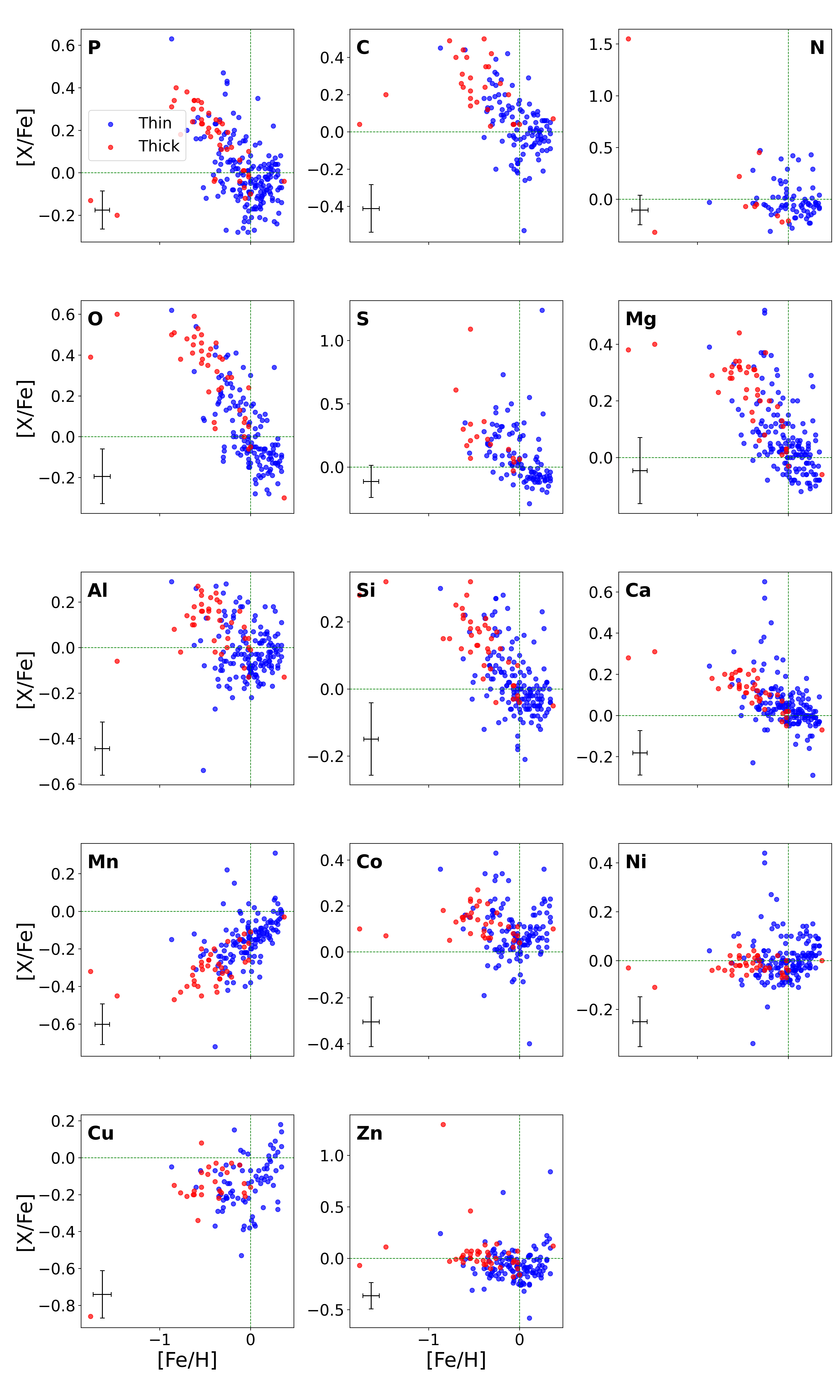}
    \caption{
    Multi-panel diagram of [X/Fe] versus [Fe/H] for the
    same elements and disk components as in
    Figure~\ref{fig:xfe_pfe_panels}. Blue and red symbols
    represent thin- and thick-disk stars, respectively;
    dashed lines indicate the solar values. A typical
    1$\sigma$ error bar in [Fe/H] and [X/Fe] is shown in the
    lower-right corner of each panel.
    }
    \label{fig:xfe_fe_panels}
\end{figure*}

In this appendix, we present the full set of multi-panel diagrams and correlation tables that complement the abundance trends discussed in Sect.~3.2. Figure~\ref{fig:xfe_pfe_panels} displays [X/Fe] as a function of [P/Fe] for all elements considered in this work, while Figure~\ref{fig:xfe_fe_panels} shows the corresponding [X/Fe]--[Fe/H] relations. For each panel, we show a typical 1$\sigma$ error bar obtained by propagating the measurement uncertainties in [X/H] and [Fe/H].

Tables~\ref{tab:xfe_pfe_corr} and \ref{tab:xfe_fe_corr} list the Spearman rank coefficients for these relations in the thin and thick disks, together with bootstrap uncertainties in the correlation strength. In the tables, $\Delta\rho^{\mathrm{boot}}$ denotes half the width of the 95\% bootstrap confidence interval for the Spearman coefficient. These diagnostic plots and tables provide the quantitative basis for the co-enrichment trends highlighted in the main text. As an additional consistency check, we repeated the main [X/H]--[P/H] correlation analysis after applying, where available, the same abundance-uncertainty cut (\(\mathrm{err}<0.15\) dex) to the other elements as that used for P and Fe. Table~\ref{tab:qualitycut_elements} compares the resulting correlation coefficients and bootstrap slopes for the full and error-restricted samples. For C, O, Mg, and Si, the results remain stable in both disk populations, whereas the thick-disk results for N and S remain less well constrained owing to the smaller sample sizes.

\begin{table*}
    \centering
        \caption{Spearman correlations between [X/Fe] and [P/Fe].}
    \label{tab:xfe_pfe_corr}
    \begin{tabular}{lrrrrrrrr}
\toprule
Element & $N_{\mathrm{thin}}$ & $\rho_{\mathrm{thin}}$ & $\Delta\rho_{\mathrm{thin}}^{\mathrm{boot}}$ & $p_{\mathrm{thin}}$ & $N_{\mathrm{thick}}$ & $\rho_{\mathrm{thick}}$ & $\Delta\rho_{\mathrm{thick}}^{\mathrm{boot}}$ & $p_{\mathrm{thick}}$ \\
\midrule
P   & 171 &  1.00 & 0.00 & $0$                   &  44 &  1.00 & 0.00 & $0$                   \\
C   & 111 &  0.49 & 0.16 & $5.6\times10^{-8}$    &  27 &  0.51 & 0.32 & $6.3\times10^{-3}$    \\
N   &  73 &  0.11 & 0.23 & $3.5\times10^{-1}$    &  10 &  0.42 & 0.71 & $2.3\times10^{-1}$    \\
O   & 133 &  0.42 & 0.15 & $4.6\times10^{-7}$    &  39 &  0.71 & 0.27 & $5.2\times10^{-7}$    \\
S   & 102 &  0.35 & 0.19 & $3.8\times10^{-4}$    &  16 &  0.63 & 0.38 & $9.6\times10^{-3}$    \\
Mg  & 155 &  0.42 & 0.15 & $7.1\times10^{-8}$    &  40 &  0.55 & 0.32 & $2.6\times10^{-4}$    \\
Al  & 148 &  0.28 & 0.17 & $6.5\times10^{-4}$    &  39 &  0.69 & 0.16 & $9.5\times10^{-7}$    \\
Si  & 155 &  0.45 & 0.15 & $4.3\times10^{-9}$    &  40 &  0.59 & 0.32 & $7.3\times10^{-5}$    \\
Ca  & 155 &  0.17 & 0.18 & $3.2\times10^{-2}$    &  40 &  0.53 & 0.33 & $4.6\times10^{-4}$    \\
Mn  & 133 & -0.09 & 0.18 & $2.9\times10^{-1}$    &  39 & -0.46 & 0.29 & $2.9\times10^{-3}$    \\
Co  & 122 &  0.34 & 0.18 & $1.3\times10^{-4}$    &  36 &  0.62 & 0.19 & $5.7\times10^{-5}$    \\
Ni  & 155 &  0.30 & 0.15 & $1.2\times10^{-4}$    &  40 &  0.20 & 0.34 & $2.1\times10^{-1}$    \\
Cu  &  76 &  0.40 & 0.21 & $3.9\times10^{-4}$    &  27 & -0.05 & 0.43 & $8.2\times10^{-1}$    \\
Zn  & 123 &  0.44 & 0.15 & $2.6\times10^{-7}$    &  40 &  0.35 & 0.30 & $2.7\times10^{-2}$    \\
\bottomrule
\end{tabular}
    \tablefoot{For each element we list the number of stars ($N_{\mathrm{thin}}$, $N_{\mathrm{thick}}$), the correlation coefficients ($\rho_{\mathrm{thin}}$, $\rho_{\mathrm{thick}}$), their bootstrap uncertainties ($\Delta\rho^{\mathrm{boot}}$), and the corresponding $p$ values. The thick-disk sample follows the original Hypatia classification and therefore includes HD~160617 and HD~211998, although these stars are identified as halo objects in the literature.}
\end{table*}

\begin{table*}
    \centering
        \caption{Spearman correlations between [X/Fe] and [Fe/H].}
    \label{tab:xfe_fe_corr}
    \begin{tabular}{lrrrrrrrr}
\toprule
Element & $N_{\mathrm{thin}}$ & $\rho_{\mathrm{thin}}$ & $\Delta\rho_{\mathrm{thin}}^{\mathrm{boot}}$ & $p_{\mathrm{thin}}$ & $N_{\mathrm{thick}}$ & $\rho_{\mathrm{thick}}$ & $\Delta\rho_{\mathrm{thick}}^{\mathrm{boot}}$ & $p_{\mathrm{thick}}$ \\
\midrule
P   & 171 & -0.37 & 0.14 & $9.0\times10^{-7}$  &  44 & -0.60 & 0.31 & $1.9\times10^{-5}$ \\
C   & 111 & -0.47 & 0.15 & $1.7\times10^{-7}$  &  27 & -0.40 & 0.37 & $3.6\times10^{-2}$ \\
N   &  73 & -0.14 & 0.21 & $2.2\times10^{-1}$  &  10 & -0.38 & 0.72 & $2.8\times10^{-1}$ \\
O   & 133 & -0.66 & 0.10 & $3.5\times10^{-18}$ &  39 & -0.81 & 0.12 & $3.4\times10^{-10}$ \\
S   & 102 & -0.58 & 0.13 & $1.5\times10^{-10}$ &  16 & -0.71 & 0.31 & $2.0\times10^{-3}$ \\
Mg  & 155 & -0.56 & 0.12 & $2.5\times10^{-14}$ &  40 & -0.74 & 0.19 & $5.1\times10^{-8}$ \\
Al  & 148 & -0.02 & 0.17 & $7.9\times10^{-1}$  &  39 & -0.40 & 0.31 & $1.3\times10^{-2}$ \\
Si  & 155 & -0.47 & 0.13 & $7.3\times10^{-10}$ &  40 & -0.78 & 0.15 & $4.2\times10^{-9}$ \\
Ca  & 155 & -0.49 & 0.13 & $1.1\times10^{-10}$ &  40 & -0.81 & 0.14 & $2.1\times10^{-10}$ \\
Mn  & 133 &  0.58 & 0.13 & $3.2\times10^{-13}$ &  39 &  0.63 & 0.22 & $1.6\times10^{-5}$ \\
Co  & 122 & -0.04 & 0.20 & $6.4\times10^{-1}$  &  36 & -0.40 & 0.28 & $1.7\times10^{-2}$ \\
Ni  & 155 &  0.21 & 0.16 & $9.4\times10^{-3}$  &  40 &  0.00 & 0.34 & $9.9\times10^{-1}$ \\
Cu  &  76 &  0.38 & 0.21 & $8.1\times10^{-4}$  &  27 &  0.33 & 0.35 & $9.3\times10^{-2}$ \\
Zn  & 123 &  0.01 & 0.19 & $8.8\times10^{-1}$  &  40 & -0.25 & 0.33 & $1.2\times10^{-1}$ \\
\bottomrule
\end{tabular}
    \tablefoot{For each element we list the number of stars ($N_{\mathrm{thin}}$, $N_{\mathrm{thick}}$), the correlation coefficients ($\rho_{\mathrm{thin}}$, $\rho_{\mathrm{thick}}$), their bootstrap uncertainties ($\Delta\rho^{\mathrm{boot}}$), and the corresponding $p$ values. The thick-disk sample follows the original Hypatia classification and therefore includes HD~160617 and HD~211998, although these stars are identified as halo objects in the literature.}
\end{table*}

\begin{table*}
\centering
\footnotesize
\setlength{\tabcolsep}{4pt}
\caption{Effect of the abundance-uncertainty cut on the main [X/H]--[P/H] relations.}
\label{tab:qualitycut_elements}
\begin{tabular}{l l l c c c c}
\toprule
Element & Sample & Population & $N$ & $\rho$ & Slope & 95\% CI \\
\midrule
C  & $\mathrm{err}<0.15$ & thick & 18  & 0.856 & 1.116 & [0.623, 2.133] \\
C  & $\mathrm{err}<0.15$ & thin  & 106 & 0.786 & 0.578 & [0.405, 0.804] \\
C  & full       & thick & 27  & 0.732 & 0.930 & [0.724, 1.163] \\
C  & full       & thin  & 111 & 0.766 & 0.577 & [0.402, 0.801] \\
N  & $\mathrm{err}<0.15$ & thick & 8   & 0.443 & 0.511 & [-2.108, 1.575] \\
N  & $\mathrm{err}<0.15$ & thin  & 68  & 0.677 & 0.847 & [0.642, 1.089] \\
N  & full       & thick & 10  & 0.262 & 0.388 & [-0.259, 1.066] \\
N  & full       & thin  & 73  & 0.683 & 0.852 & [0.641, 1.094] \\
O  & $\mathrm{err}<0.15$ & thick & 23  & 0.883 & 0.459 & [0.284, 0.971] \\
O  & $\mathrm{err}<0.15$ & thin  & 113 & 0.665 & 0.556 & [0.385, 0.702] \\
O  & full       & thick & 39  & 0.840 & 0.568 & [0.381, 0.805] \\
O  & full       & thin  & 133 & 0.643 & 0.551 & [0.384, 0.691] \\
S  & $\mathrm{err}<0.15$ & thick & 14  & 0.490 & 0.832 & [-0.807, 2.000] \\
S  & $\mathrm{err}<0.15$ & thin  & 90  & 0.618 & 0.577 & [0.378, 0.694] \\
S  & full       & thick & 16  & 0.428 & 0.340 & [-1.011, 1.691] \\
S  & full       & thin  & 102 & 0.622 & 0.576 & [0.377, 0.693] \\
Mg & $\mathrm{err}<0.15$ & thick & 35  & 0.883 & 0.857 & [0.640, 1.026] \\
Mg & $\mathrm{err}<0.15$ & thin  & 139 & 0.817 & 0.734 & [0.607, 0.886] \\
Mg & full       & thick & 40  & 0.895 & 0.849 & [0.633, 1.011] \\
Mg & full       & thin  & 155 & 0.832 & 0.732 & [0.608, 0.883] \\
Si & $\mathrm{err}<0.15$ & thick & 37  & 0.916 & 1.018 & [0.752, 1.219] \\
Si & $\mathrm{err}<0.15$ & thin  & 146 & 0.839 & 0.730 & [0.641, 0.822] \\
Si & full       & thick & 40  & 0.919 & 1.003 & [0.748, 1.191] \\
Si & full       & thin  & 155 & 0.835 & 0.731 & [0.639, 0.823] \\
\bottomrule
\end{tabular}
\tablefoot{The error-restricted sample applies, where available, the same abundance-uncertainty cut ($\mathrm{err}<0.15$ dex) to the additional elements as that used for P and Fe. We report the sample size, the Spearman rank coefficient, and the bootstrap median slope with its 95\% confidence interval.}
\end{table*}

\section{Additional analysis on molar ratios}
\label{app:CO_MgSi}

As discussed in Sect.~\ref{sec:molratio}, the C/O and Mg/Si molar ratios are classical diagnostics of bulk planetary composition. As discussed by \citet{thiabaud2015elemental}, C/O determines whether refractory condensates in the protoplanetary disk are dominated by silicates or whether carbon-rich phases, such as carbides and graphite, can become important, whereas Mg/Si controls the relative proportions of olivine- and pyroxene-type silicates in mantle mineralogy. For completeness, we therefore examined the distributions of C/O and Mg/Si in our P-star sample and compared them with those of the control sample, which traces the local FGK field population.

\begin{figure*}
    \centering
    \includegraphics[width=0.48\textwidth]{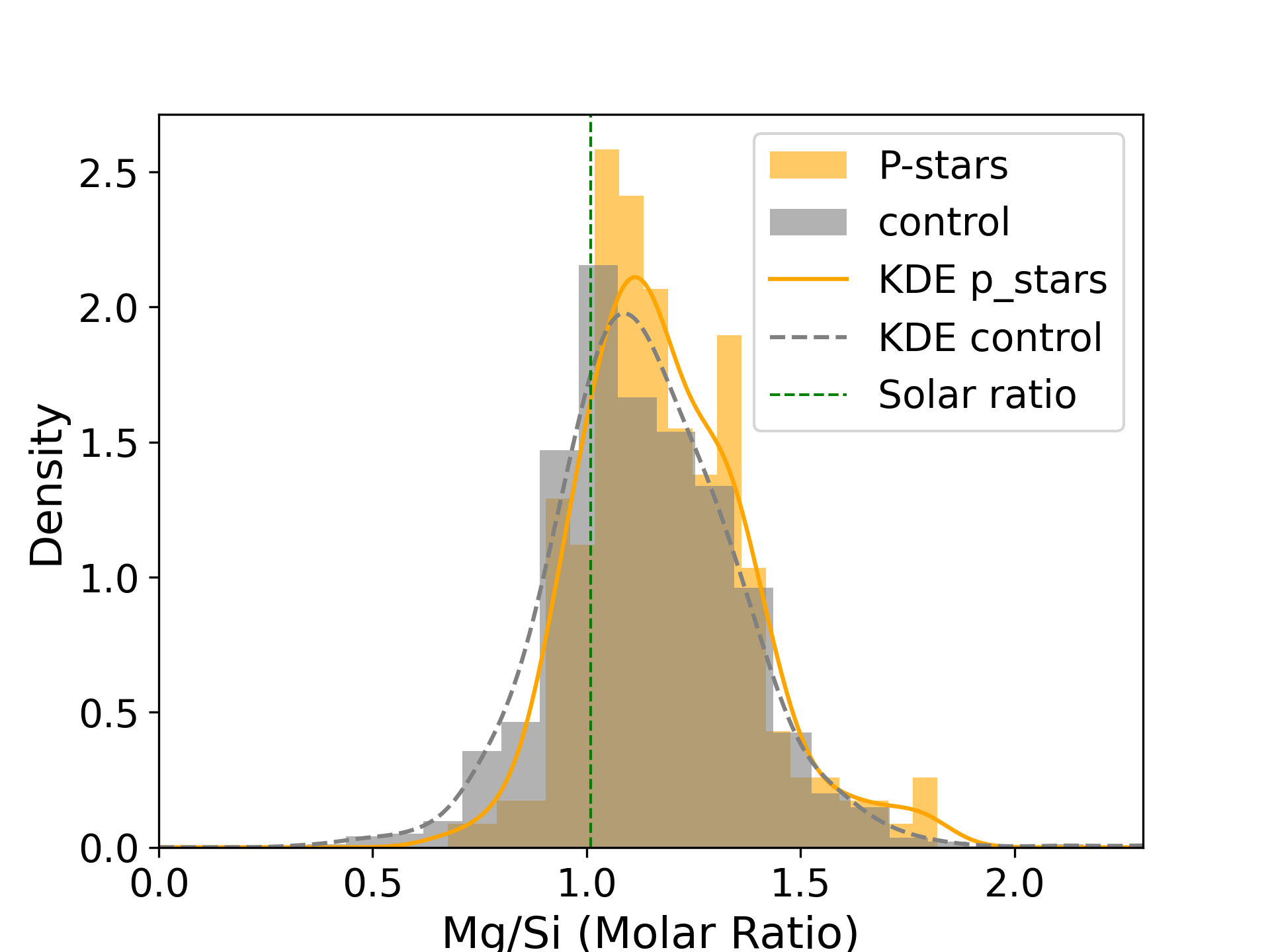}
    \hfill
    \includegraphics[width=0.48\textwidth]{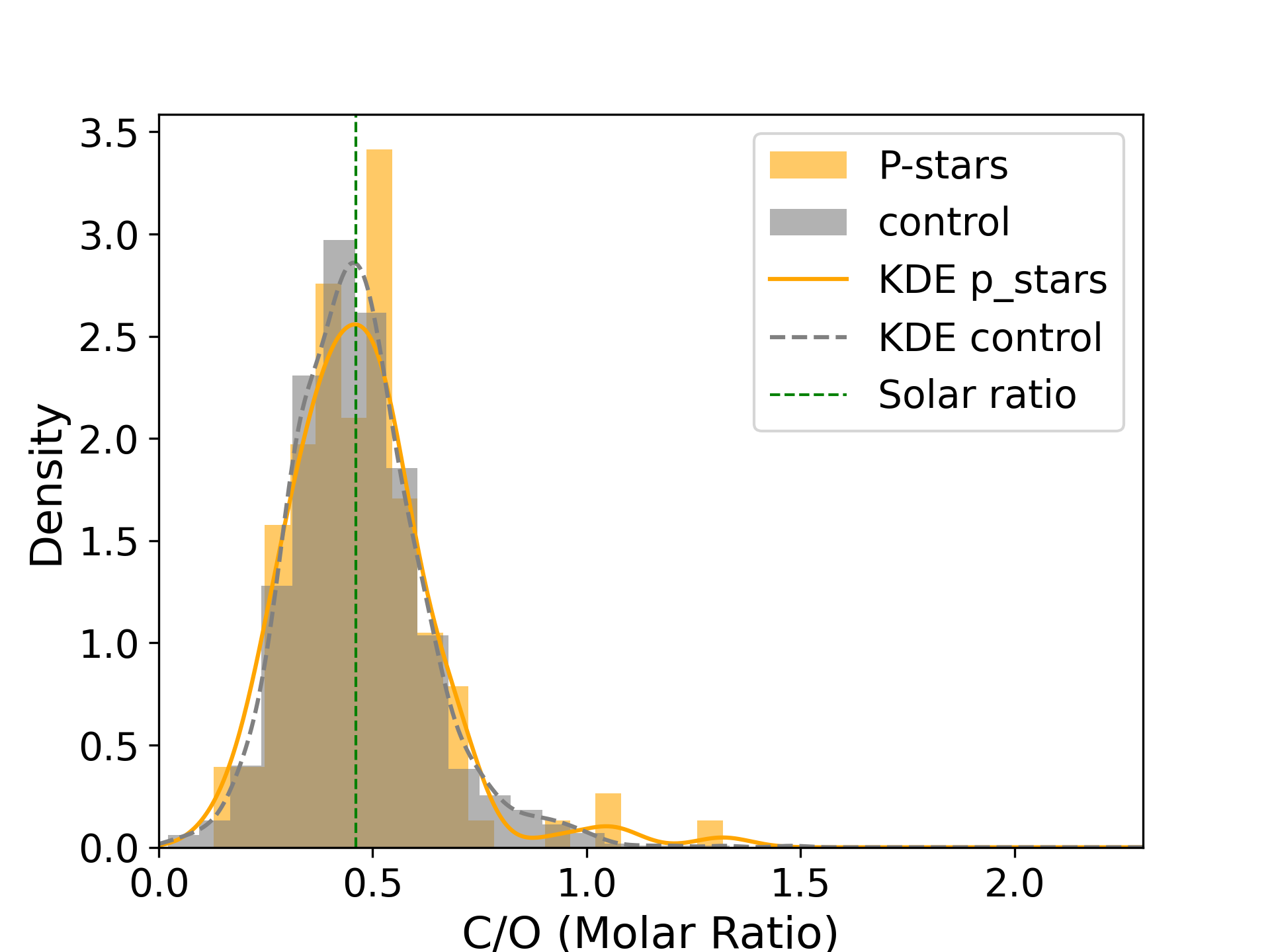}
    \caption{Distributions of Mg/Si and C/O for the P-star and control samples. Histograms and kernel density estimates are shown for both samples. The green dotted line marks the solar photospheric value of each ratio.}
    \label{fig:mgsidistr}
\end{figure*}

Figure~\ref{fig:mgsidistr} shows the resulting histograms for both ratios. The C/O distributions of the P-star and control samples are broadly similar, indicating that our P detections are not confined to an extreme tail of the C/O distribution. The Mg/Si distributions differ only modestly, with the P-star sample showing, at most, a mild shift towards higher Mg/Si compared with the control sample. This check confirms that the stars in which P has been measured are not a chemically peculiar subset in terms of the classical C/O--Mg/Si parameter space and that the multi-element co-enrichment patterns discussed in the main text are not driven by a strong selection in these two mineralogy-sensitive ratios.

\begin{figure*}
  \centering
  \includegraphics[width=0.8\textwidth]{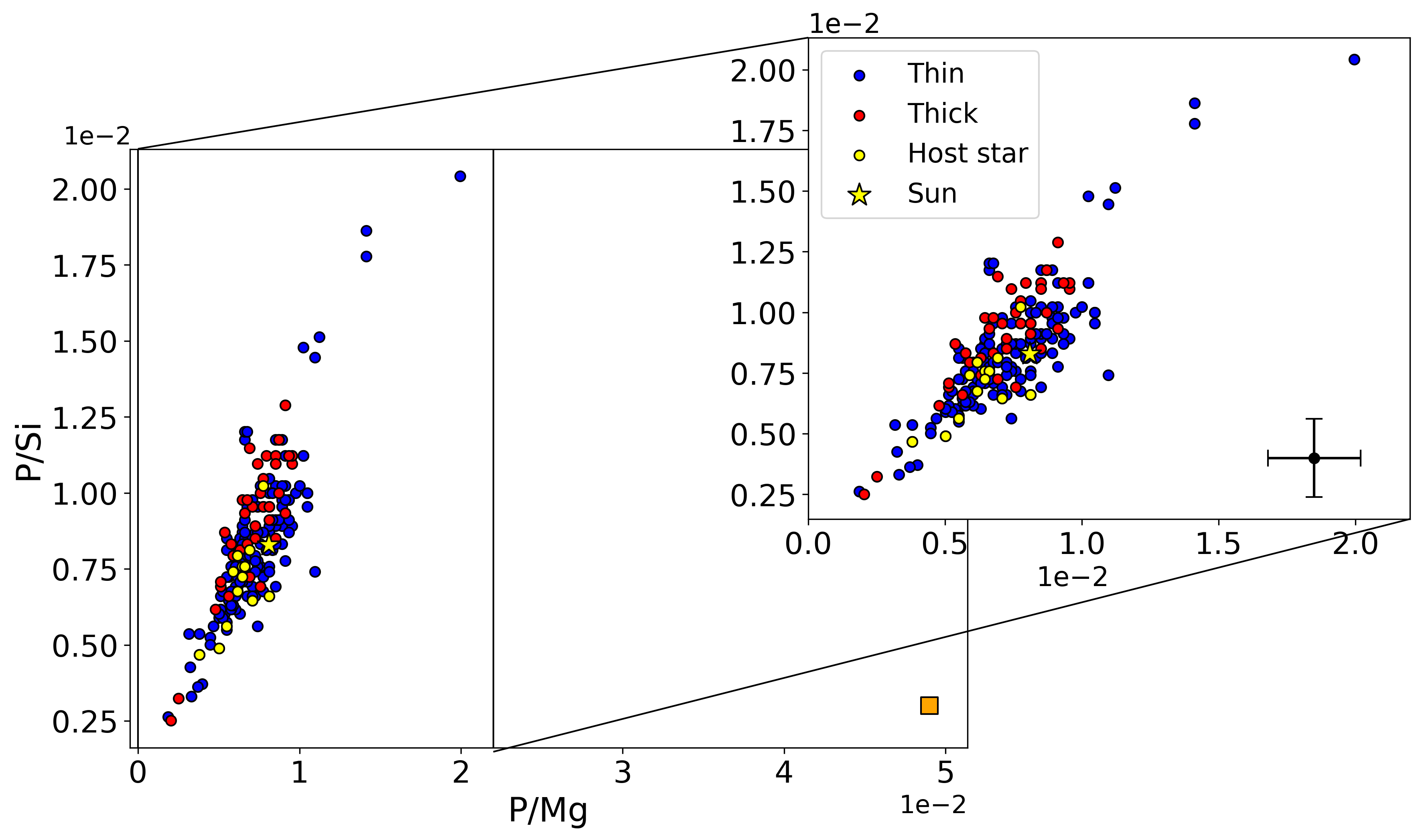}
  \caption{P/Si as a function of P/Mg. The enlarged panel highlights the region containing the highest density of data points. The solar and terrestrial upper-crust values are marked by a yellow star and an orange square, respectively. The average uncertainties are shown in the lower-right corner of the enlarged panel. Both thin- and thick-disk stars show a very tight correlation between P/Si and P/Mg. Part of this behaviour is expected on algebraic grounds because, in logarithmic space, $\log(\mathrm{P/Si}) - \log(\mathrm{P/Mg}) = \log(\mathrm{Mg/Si})$, so the scatter is largely driven by the Mg/Si distribution. This diagram therefore illustrates how the bulk P/Mg and P/Si ratios co-vary coherently with Mg/Si and with the overall $\alpha$-element enrichment of disk stars, constraining the initial chemical inventory available to rocky planets rather than directly probing planetary differentiation or P partitioning.}

  \label{fig:siliciomagnesio}
\end{figure*}

Figure~\ref{fig:siliciomagnesio} shows the relation between the molar ratios P/Mg and P/Si for thin- and thick-disk stars, together with the positions of the Sun and Earth's upper crust. Both disk populations show a very tight correlation between P/Si and P/Mg. Part of this tightness is expected on algebraic grounds because both ratios share P in the numerator and, in logarithmic space,
$\log(\mathrm{P/Si}) - \log(\mathrm{P/Mg}) = \log(\mathrm{Mg/Si})$.
A simple check confirms that the scatter around the P/Si--P/Mg relation is entirely accounted for by the observed Mg/Si distribution: in both disks, $\log(\mathrm{P/Si}) - \log(\mathrm{P/Mg})$ and $\log(\mathrm{Mg/Si})$ have virtually identical dispersions and a Spearman rank coefficient of $\rho_{\mathrm{S}}\simeq 1.0$. We therefore interpret Figure~\ref{fig:siliciomagnesio} as showing that the initial bulk P/Mg and P/Si ratios co-vary coherently with the Mg/Si distribution and with the overall $\alpha$-element enrichment of disk stars. This relation constrains the initial chemical inventory available to rocky planets rather than directly probing planet-scale differentiation or P partitioning between the core, mantle, and crust.

The solar values shown in the figure, marked by a yellow star, lie near the centre of the distribution, indicating that the solar composition is representative in terms of these ratios. Planet-host stars do not show a distinct relation between the two ratios compared with stars without detected planets. Because Earth's crust is a key near-surface reservoir of bioavailable P \citep{walton2023phosphorus}, we compared its ratios with those inferred from stellar abundances. The molar ratios for Earth's upper crust are P/Mg = 0.049 and P/Si = 0.003, calculated from the abundances reported by \citet{yaroshevsky2006abundances}; these values differ substantially from the stellar distribution. This difference highlights how strongly differentiation processes during planet formation and evolution may affect elemental ratios and planetary composition.

\end{appendix}

\end{document}